\documentclass[fleqn,usenatbib]{mnras}

\usepackage{newtxtext,newtxmath}
\usepackage[T1]{fontenc}

\DeclareRobustCommand{\VAN}[3]{#2}
\let\VANthebibliography\thebibliography
\def\thebibliography{\DeclareRobustCommand{\VAN}[3]{##3}\VANthebibliography}

\usepackage{xspace}
\usepackage{xcolor}
\definecolor{darkblue}{rgb}{0.0, 0.0, 0.55}
\definecolor{darkgreen}{rgb}{0.0, 0.55, 0.2}
\definecolor{darkred}{rgb}{0.55, 0.0, 0}
\usepackage{multirow}
\usepackage{mathtools}
\usepackage{bm}
\usepackage{comment}
\usepackage{graphicx}	% Including figure files
\usepackage{amsmath}	% Advanced maths commands
\newcommand{\pto}{$\Hat{P}_{21}$\xspace}
\newcommand{\bias}{b_{\rm HI}\xspace}

\newcommand{\secref}[1]{\hyperref[#1]{section~\ref*{#1}}}
\newcommand{\appref}[1]{\hyperref[#1]{appendix~\ref*{#1}}}

\DeclareMathOperator{\diag}{diag}

\title[Forecasted cosmological parameters estimation for the SKA Observatory]{Multipole expansion for 21cm Intensity Mapping power spectrum: forecasted cosmological parameters estimation for the SKA Observatory}

\author[M. Berti et al.]{
Maria Berti,$^{1,2,3}$\thanks{E-mail: mberti@sissa.it}
Marta Spinelli,$^{4,5,6}$
and Matteo Viel$^{1,2,3,5}$
\\
$^{1}$SISSA- International School for Advanced Studies, Via Bonomea 265, 34136 Trieste, Italy\\
$^{2}$INFN – National Institute for Nuclear Physics, Via Valerio 2, I-34127 Trieste, Italy\\
$^{3}$IFPU, Institute for Fundamental Physics of the Universe, via Beirut 2, 34151 Trieste, Italy\\
$^{4}$Institute for Particle Physics and Astrophysics,
ETH Z{\"u}rich, Wolfgang Pauli Strasse 27, 8093 Z{\"u}rich, Switzerland\\
$^{5}$INAF, Osservatorio Astronomico di Trieste, Via G. B. Tiepolo 11, I-34131 Trieste, Italy\\
$^{6}$Department of Physics and Astronomy, University of the Western Cape, Robert Sobukhwe Road, Bellville, 7535, South Africa\\
}

\date{Accepted XXX. Received YYY; in original form ZZZ}

\pubyear{2022}

\begin{document}
\label{firstpage}
\pagerange{\pageref{firstpage}--\pageref{lastpage}}
\maketitle

% Abstract of the paper MAX 250 parole
\begin{abstract}
The measurement of the large scale distribution of neutral hydrogen in the late Universe, obtained with radio telescopes through the hydrogen 21cm line emission, has the potential to become a key cosmological probe in the upcoming years.
We explore the constraining power of 21cm intensity mapping observations on the full set of cosmological parameters that describe the $\Lambda$CDM model. We assume a single-dish survey for the SKA Observatory and simulate the 21cm linear power spectrum monopole and quadrupole within six redshift bins in the range $z=0.25-3$. Forecasted constraints are computed numerically through Markov Chain Monte Carlo techniques. We extend the sampler \texttt{CosmoMC} by implementing the likelihood function for the 21cm power spectrum multipoles. We assess the constraining power of the mock data set alone and combined with Planck 2018 CMB observations. We include a discussion on the impact of extending measurements to non-linear scales in our analysis. We find that 21cm multipoles observations alone are enough to obtain constraints on the cosmological parameters comparable with other probes. Combining the 21cm data set with CMB observations results in significantly reduced errors on all the cosmological parameters. The strongest effect is on $\Omega_ch^2$ and $H_0$, for which the error is reduced by almost a factor four. The percentage errors we estimate are $\sigma_{\Omega_ch^2} = 0.25\%$ and $\sigma_{H_0} = 0.16\%$, to be compared with the Planck only results $\sigma_{\Omega_ch^2} = 0.99\%$ and $\sigma_{H_0} = 0.79\%$. We conclude that 21cm SKAO observations will provide a competitive cosmological probe, complementary to CMB and, thus, pivotal for gaining statistical significance on the cosmological parameters constraints, allowing a stress test for the current cosmological model.
\end{abstract}


%%%%%%%%%%%%%%%%%%%%%%%%%%%%%%%%%%%%%%%%%%%%%%%%%%

%%%%%%%%%%%%%%%%% BODY OF PAPER %%%%%%%%%%%%%%%%%%

\section{Introduction}

Neutral hydrogen (HI) is a fundamental element in the Universe and its late-time distribution traces the underlying matter field, making it an innovative key probe of the large scale structure (LSS) \citep[e.g.][]{review,Ansari_2012,Santos:2015}. 
Despite the success of Cosmic Microwave Background (CMB) experiments~\citep{WMAP,planck:2018,ACT} and the LSS measurement via galaxy surveys \citep[e.g.][]{BOSS:2016} in constraining the cosmological parameters of the $\Lambda$CDM model, we still lack an understanding of the nature of dark energy and dark matter and an explanation for some of the tensions among different observables~\citep[e.g.][]{riess:2019,wong:2019,Verde:2019}. The measurement of the large scale distribution of HI and its evolution with time can thus play an important role in the upcoming years, providing a complementary probe to traditional galaxy surveys~\citep[e.g.][]{bull:2015}. 

The 21cm signal, originating from the spin-flip transition in the hyperfine structure of the hydrogen ground state (e.g.~\citealt{Furlanetto:2006}), is redshifted by the expansion of the Universe, and, thus, it is detectable on Earth at radio frequencies. Several planned and ongoing experiments, either purpose-built compact interferometers (such as CHIME~\citep{Bandura:2014gwa,CHIMEdetection}, CHORD or HIRAX~\citep{Newburgh:2016mwi}) or single-dish telescopes (such as GBT~\citep{Masui2013,Wolz2022} or FAST~\citep{Hu:2019okh}) aim to measure it with intensity mapping (IM) techniques~\citep{Bharadwaj:2000,Battye:2004,McQuinn:2005,Chang:2007,Seo:2009fq,Kovetz:2017,Villaescusa-Navarro:2018} and some of them have achieved the detection of the HI signal in cross correlation with galaxy surveys \citep{Chang2010,Masui2013,Anderson2018,Wolz2022}.

Radio cosmology is also one of the main science goals of the SKA Observatory (SKAO)\footnote{\url{https://www.skao.int/}}, that will be composed by SKA-Low and
SKA-Mid telescopes located in Australia and South Africa, respectively. Using the SKA-Mid telescope array as a collection of single-dishes~\citep{Battye:2013,Santos:2015} it will be possible to perform 21cm intensity mapping observations at the large scale important for cosmology up to redshift 3~\citep{Bacon:2018}.  
The SKAO is currently under construction, and MeerKAT, the SKA-Mid precursor, has been conducting IM survey for cosmology~\citep[MeerKLASS]{Santos:2017}.
Preliminary data analysis have provided promising results~\citep{Wang:2021,irfan2022} and a first detection of the HI signal in cross correlation with the WiggleZ galaxies~\citep{Cunnington:2022uzo}. However, the level of foreground residuals is preventing a direct detection and this issue has triggered an extensive simulation work on foreground cleaning performances~\citep{Alonso2015,Wolz2016,Carucci2020,Matshawule:2021,Irfan2021, Cunnington2021,Soares:2020,Soares2021GPR,Spinelli:2021emp}.
In parallel with the effort in improving the data analysis and the foreground separation, it is of key importance to refine the forecast for the constraining power of the 21cm IM alone and in combination with other probes in order to make a better case for radio cosmology with the SKAO or optimize the survey design.

\medskip
In this work, we focus on the parameters of the $\Lambda$CDM model taking into account the redshift-space nature of the 21cm power spectrum. In this way we can exploit the tomographic, i.e. refined in redshift, potential of the observations. We build on the formalism of~\citet{Blake:2019,Cunnington:2020, Soares:2020} and study redshift-space power spectrum monopole and quadrupole. Following \citet{Bacon:2018}, 
we construct mock 21cm power spectrum measurements in six redshift bins, in the redshift range $z=0.25 - 3$.
We expand the code \texttt{CosmoMC}~\citep{Lewis:2002,Lewis:2013} to include a new likelihood module to compute constraints through Markov-Chain Monte-Carlo (MCMC) techniques and we assess the constraining power of our mock 21cm data set alone and combined with CMB data. 

We note that forecasts for future IM observations based on the Fisher Matrix formalism, thereby using a complementary approach to the one described here, have been presented in \citet{obuljen18}.
 
The novelty aspects of this study with respect to \citet{Soares:2020}, are: $i)$ we produce constraints on the full set of cosmological parameters; $ii)$ we combine the forecasted 21cm power spectrum multipoles with Planck 2018 CMB data~\citep{planck:2018} ; $iii)$ we employ the multipole formalism to construct and study the constraining power of a tomographic data set. 
We neglect the effect of foregrounds but we include uncertainties on the astrophysical quantities that connect the measured 21cm power spectrum to the underlying matter field, as the atomic hydrogen bias and the brightness temperature (see also \citealt{Berti:2021}).

\medskip
The structure of the paper is the following. The adopted methodology, including the modelling of the 21cm power spectrum and multipoles, the construction of the forecasted data set and the likelihood implementation, is discussed in \autoref{sec:methods}. Results are presented in \secref{sec:results}. The constraining power of the mock 21cm power spectrum multipoles is evaluated in \secref{sec:res_P21}. We investigate how 21cm observations affect the constraining power of other probes, i.e. CMB measurements, in \secref{sec:res_P21_P21+Planck}. A discussion on the impact of opening the parameter space to the brightness temperature, the HI bias and the growth rate is given in \secref{sec:res:nuisances}. We also investigate the extension to non-linear scale of our mock data in \secref{sec:res_non_linear}. A summary of the results and our conclusions are outlined in \secref{sec:conclusions}.

\medskip
Through all this work we assume a $\Lambda$CDM universe described by a Planck 2018~\citep{planck:2018}, fiducial cosmology i.e. $\{\Omega_b h^2 = 0.022383,\, \Omega_c h^2 = 0.12011,\, n_s= 0.96605 ,\, \ln (10^{10} A_s) =3.0448 ,\, \tau= 0.0543,\, h = 0.6732,\, \Sigma m_\nu = 0.06 \mathrm{eV} \}$.

\section{Methods}
\label{sec:methods}

In this section we outline the formalism and the methods used 
throughout this work. 
We describe the 21cm linear power spectrum from \secref{sec:P21_model} to~\secref{sec:mult_exp}. In \secref{sec:errors} and~\secref{sec:tomographic_data_set} we construct the mock tomographic data set of SKA-Mid observations. The description of the likelihood function and the parameter estimation method are reported in \secref{sec:bayes}.

\subsection{The theoretical 21cm signal linear power spectrum}
\label{sec:P21_model}

We model the 21cm linear power spectrum as~\citep{kaiser1987,Villaescusa-Navarro:2018, Bacon:2018}
\begin{equation}\label{eq:P21}
     P_{21}(z,\,k,\, \mu) = \Bar{T}_{\rm b}^2(z) \left[ b_{\mathrm{HI}}(z) + f(z)\, \mu^2\right]^2 P_{\rm m}(z,k),
\end{equation}
where $\bar{T}_{\rm b}$ is the HI mean brightness temperature, $\bias$ is the HI bias, $f$ is the growth rate, $\mu= \hat{k} \cdot \hat{z}$ is the cosine of the angle between the wave number and the line-of-sight, and $P_{\rm m}(z,k)$ is the linear matter power spectrum. 
We neglect in \autoref{eq:P21} the shot noise term, which is believed to be negligible at linear scales~\citep{villa14,Pourtsidou:2016,Villaescusa-Navarro:2018,Spinelli:2019}. We refer to \secref{sec:res_non_linear} for a discussion on non-linear scales and the shot noise term.

We use the parametrization of the brightness temperature from~\cite{Battye:2013} 
\begin{equation}\label{eq:Tb}
    \bar{T}_{\mathrm{b}} (z) = 180 \, \,  \Omega_{\mathrm{HI}}(z) \frac{h\, H_0}{ H(z)} (1+z)^2 \text{mK},
\end{equation}
where we consider the HI density parameter to evolve mildly in redshift as $\Omega_{\mathrm{HI}}(z) = 4. \times 10^{-4} (1+z)^{0.6} $~\citep[see][]{Crighton:2015}. Given that we lack an analytical model, $\bias (z)$ at given redshift is computed by interpolating numerical results from hydro-dynamical simulations~\citep{Villaescusa-Navarro:2018,villa15}.

The growth rate $f(z)$ and the linear matter power spectrum $P_{\rm m}(z,k)$ are, instead, computed numerically by means of the Einstein-Boltzmann solver \texttt{CAMB}\footnote{See \url{https://camb.info/}.}~\citep{Lewis:2000}.

\subsection{Modeling the observed 21cm signal power spectrum as measured by SKA-Mid}
\label{sec:beam_model}
\begin{table}
	\centering
	\caption{Assumed specifications for SKA-Mid survey~\citep{Bacon:2018}.}
	\label{tab:SKA_specifics}
	\begin{tabular*}{\columnwidth}{l@{\hspace*{40pt}}l@{\hspace*{40pt}}c} 
		\hline
		Parameter & & Value\\
		\hline
		$D_{\rm dish}\, [\rm m]$ & SKAO dish diameter & 15 \\
		$N_{\rm dish}$ & SKAO dishes & 133 \\
		$t_{\rm obs}\, [\rm h]$ & observing time & 10000 \\
		$T_{\rm sys}\, [\rm K]$ & system temperature & 25 \\
		$\delta \nu\, [\rm MHz]$ & frequency range & 1 \\
		$A_2 \, [\rm deg^2]$ & survey area (Band 2) & 5000 \\
		$\Omega_{\rm sur,1} \, [\rm sr] $ & survey area (Band 2) & 1.5 \\
		$A_1 \, [\rm deg^2]$ & survey area (Band 1) & 20000 \\
		$\Omega_{\rm sur,2} \, [\rm sr] $ & survey area (Band 2) & 6.1 \\
		$f_{\rm sky,2}$ & covered sky area (Band 2) & 0.12\\
		$f_{\rm sky,1}$ & covered sky area (Band 1) & 0.48\\
		$\Delta z$ & width of the redshift bins& 0.5 \\
		\hline
	\end{tabular*}
\end{table}

In this work, we construct mock single-dish 21cm power spectrum observations of the SKA-Mid telescope, modeling the 21cm IM survey as in \citet{Bacon:2018}

In the following, we discuss the effect of the telescope beam and of the instrumental noise on the power spectrum. The telescope specifications relevant for our work are reported in \autoref{tab:SKA_specifics}. 

\subsubsection{The effect of the telescope beam}
One of the main instrumental contamination is the effect of a Gaussian telescope beam, which suppresses the power spectrum on scales smaller than the beam full width at half maximum \citep{Cunnington:2022,Soares:2020,Cunnington:2020,Villaescusa-Navarro:2016, Battye:2013}.

Its effect can be written in terms of $R_{\rm beam}$, the beam physical dimension
\begin{align}
\begin{split}
 R_{\rm beam} (z) &= \sigma_\theta r(z) \\
 &= \frac{\theta_{\rm FWHM}}{2 \sqrt{2\ln 2}} r(z),
 \end{split}
\end{align}
where $r(z)$ is the comoving distance, $\theta_{\rm FWHM} = \frac{1.22 \lambda_{21}}{D_{\rm dish}} (1+z) $ is the full width at half maximum, and $D_{\rm dish}$ is the diameter of the dish. 

The beam damping factor in Fourier space  $\tilde{B} (z,k,\mu)$ can thus be written as
\begin{equation}
    \tilde{B} (z,k,\mu) = \exp \left[ \frac{-k^2 R_{\rm beam}^2(z) (1-\mu^2)}{2}\right].
\end{equation} 

Note that the factor $(1-\mu)$ model the smoothing only along the transverse direction since the damping along the radial direction is negligible due to the high frequency resolution of 21cm observation~\citep{Villaescusa-Navarro:2016}. 

The beam convolved 21cm power spectrum is then
\begin{equation}
    \label{eq:P_21_full}
     \Hat{P}_{21}(z,\,k,\, \mu) = \tilde{B}^2 (z,k,\mu) P_{21}(z,k,\mu),
\end{equation}
where $P_{21}(z,k,\mu)$ is defined in~\autoref{eq:P21}.

\subsubsection{Instrumental noise}
\label{sec:instrumental_noise}
For a single-dish intensity mapping SKAO-like experiment, the noise power spectrum can be modeled as~\citep{Santos:2015,Bernal:2019}
\begin{equation}\label{eq:noise}
    P_{\rm N}(z) = \frac{T_{\rm sys}^2 4\pi f_{\rm sky}}{N_{\rm dish} t_{\rm obs} \delta \nu} \dfrac{V_{\rm bin} (z)}{\Omega_{\rm sur}}.
\end{equation}
Here, given a redshift bin centered at $z$ and of width $\Delta z$, the volume of the redshift bin $V_{\rm bin}(z)$ can be computed as
\begin{align}\label{eq:volume_bin}
\begin{split}
       V_{\rm bin}(z) &=  \Omega_{\rm sur} \int_{z - \Delta z /2}^{z + \Delta z /2} \mathrm{d}z' \, \frac{\mathrm{d} V}{\mathrm{d} z'\mathrm{d}\Omega} \\
       &=  \Omega_{\rm sur}  \int_{z - \Delta z /2}^{z + \Delta z /2} \mathrm{d}z' \, \frac{c r(z')^2}{H(z')}.
\end{split}
\end{align}
with $r(z)$ being the comoving distance.
A description of all the other parameters that appear in \autoref{eq:noise} and their assumed values can be found in \autoref{tab:SKA_specifics}. 

\subsubsection{Variance}
To construct mock observation we need an estimation of the errors on the power spectrum. Following \citet[e.g.][]{Bernal:2019} we write the variance per $k$ and $\mu$ bin $\sigma(z,k,\mu)$ as 
\begin{equation}\label{eq:variance}
    \sigma^2(z,k,\mu) = \frac{\Big(\Hat{P}_{21}(z,k,\mu) + P_{\rm N}(z)\Big)^2}{N_{\rm modes}(z,k,\mu)},
\end{equation}
where $\Hat{P}_{21}(z,k,\mu)$ is the 21cm signal power spectrum, defined in \autoref{eq:P_21_full}, and $P_{\rm N} (z)$ is the noise power spectrum of \autoref{eq:noise}.  $N_{\rm modes}(z,k,\mu)$ is the number of modes per $k$ and $\mu$ bins in the observed sky volume. We can compute it as
\begin{equation}
\label{eq:nmodes_mu}
    N_{\rm modes}(z,k,\mu) = \frac{k^2 \Delta k(z) \Delta \mu(z)}{8\pi^2} V_{\rm bin}(z).
\end{equation}
Here, $V_{\rm bin}(z)$ is the volume of the redshift bin centered at $z$, while $\Delta k(z)$ and $\Delta \mu(z)$ are the $k$ and $\mu$ bin width respectively. In our analysis, however, we integrate over all the possible values of $\mu$ in the interval $\mu \in (-1,1)$ (as we will discuss in more detail in the next section). Thus, computing the number of $\mu$ modes, \autoref{eq:nmodes_mu} reduces to
\begin{equation}
     N_{\rm modes}(z,k) = \frac{k^2 \Delta k(z)}{4\pi^2} V_{\rm bin}(z).
\end{equation}

\subsection{Multipole expansion}
\label{sec:mult_exp}
The non isotropic redshift space 21cm power spectrum
can be decomposed using Legendre polynomials $\mathcal{L}_\ell(\mu)$ as
\begin{equation}
    \Hat{P}_{21} (z,k,\mu) = \sum_\ell \Hat{P}_\ell (z,k) \mathcal{L}_\ell(\mu).
\end{equation}
The first Legendre polynomials are the following functions of $\mu$
\begin{equation}
    \mathcal{L}_0 (\mu) = 1, \qquad \mathcal{L}_2 (\mu) = \frac{3\mu^2}{2} - \frac{1}{2}.
\end{equation}
The coefficients of the expansion, i.e. the multipoles of the 21cm power spectrum, are then given by
\begin{equation}
\label{eq:ell_P}
    \Hat{P}_\ell (z,k) = \frac{(2\ell + 1)}{2} \int_{-1}^{1} {\rm d}\mu\, \mathcal{L}_\ell(\mu) \Hat{P}_{21}(z,k,\mu),
\end{equation}
where the expression for $\Hat{P}_{21}(z,k,\mu)$ can be found in \autoref{eq:P_21_full}.

In our analysis, we construct mock observations for the monopole $\Hat{P}_0(z,k)$ and the quadrupole $\Hat{P}_2(z,k)$, i.e. $\ell = 0$ and $\ell = 2$ respectively. We refer to \appref{sec:app_equations} for the explicit analytical expression of these quantities. 

\subsection{Errors and covariance matrix}
\label{sec:errors}
To compute the errors on the monopole and the quadrupole, we begin by defining the covariance between the multipoles $\ell$ and $\ell'$ as a function of $k$ and $z$ \citep[see][]{Bernal:2019,Chung:2019}
\begin{equation}
\label{eq:Cell}
    C_{\ell\ell'}(z,k) = \frac{(2\ell + 1)(2\ell' + 1)}{2} \int_{-1}^1  {\rm d}\mu\, \mathcal{L}_\ell(\mu)\, \mathcal{L}_{\ell'}(\mu)\,\,\sigma^2(z,k,\mu),
\end{equation}
where we neglect mode coupling. Here, $\sigma^2(z,k,\mu)$ is the variance per $k$ and $\mu$ bin at redshift $z$, as defined in~\autoref{eq:variance}.

\autoref{eq:Cell} can be used both to estimate the error on a single data point and to compute the covariance matrix for the multipoles. Given a set of $N$ measurements of the 21cm multipole $\Hat{P}_\ell$ at scales $\{k_1, \dots, k_N\}$, the computed error on each point of the data set is 
\begin{align}\label{eq:errors}
\begin{split}
        \sigma_{\Hat{P}_\ell}(z,k_i) &=  \sqrt{C_{\ell\ell}(z,k_i)} \\
        &=\sqrt{\frac{(2\ell + 1)^2}{2} \int_{-1}^1  {\rm d}\mu\, \mathcal{L}^2_\ell(\mu)\,\sigma^2(z,k_i,\mu)},
\end{split}
\end{align}
for each $k_i$ in $\{k_1,\dots, k_N\}$. 
In our analysis we focus only on the monopole $\Hat{P}_0$ and the quadrupole $\Hat{P}_2$.

The multipole covariance defined in~\autoref{eq:Cell} allows to estimate both the covariance for a given multipole and between different multipoles. Thus, the most general covariance matrix for $\Hat{P}_0$ and $\Hat{P}_2$ combined is a $2N \times 2N$ symmetric block matrix. At fixed redshift, it is constructed as
\begin{equation}\label{eq:full_covariance}
    \bm{\mathsf{C}}(z) = \begin{pmatrix}
\bm{\mathsf{C}}_{00}(z) & \bm{\mathsf{C}}_{02}(z)\\
\bm{\mathsf{C}}_{02}(z) & \bm{\mathsf{C}}_{22}(z)
\end{pmatrix}.
\end{equation}
Each block $\bm{\mathsf{C}}_{\ell\ell'}$ is a diagonal matrix of dimensions $N\times N$, defined as $\bm{\mathsf{C}}_{\ell\ell'}(z) = \diag(C_{\ell\ell'}(z,k_1), \dots, C_{\ell\ell'}(z,k_N))$, where the elements $C_{\ell\ell'}(z,k_i)$ are computed as in~\autoref{eq:Cell} at each $\{k_1,\dots, k_N\}$.

The blocks along the diagonal, i.e. $\bm{\mathsf{C}}_{00}(z)$ and $\bm{\mathsf{C}}_{22}(z)$, are the covariance matrices for the monopole and the quadrupole alone. The off-diagonal block $\bm{\mathsf{C}}_{02}(z)$, instead, describes the covariance between $\Hat{P}_0$ and $\Hat{P}_2$. 

In the simplified case where the monopole $\Hat{P}_0$ and the quadrupole $\Hat{P}_2$ are uncorrelated, we can neglect the off-diagonal terms in $\bm{\mathsf{C}}(z)$ and assume the block-diagonal covariance matrix
\begin{equation}\label{eq:covariance_diagonal}
    \bm{\mathsf{C}}_{\rm diag}(z) = \begin{pmatrix}
\bm{\mathsf{C}}_{00}(z) & 0\\
0 & \bm{\mathsf{C}}_{22}(z)
\end{pmatrix}.
\end{equation}
In our work, we compute the covariance matrices $\bm{\mathsf{C}}_{00}(z)$, $\bm{\mathsf{C}}_{02}(z)$, $\bm{\mathsf{C}}_{22}(z)$, $\bm{\mathsf{C}}_{\rm diag}(z)$, and $\bm{\mathsf{C}}(z)$ numerically, from the analytical expression for the monopole and the quadrupole (see \appref{sec:app_equations}).

\subsection{Tomographic data set}
\label{sec:tomographic_data_set}
\begin{figure}
	\includegraphics[width=\columnwidth]{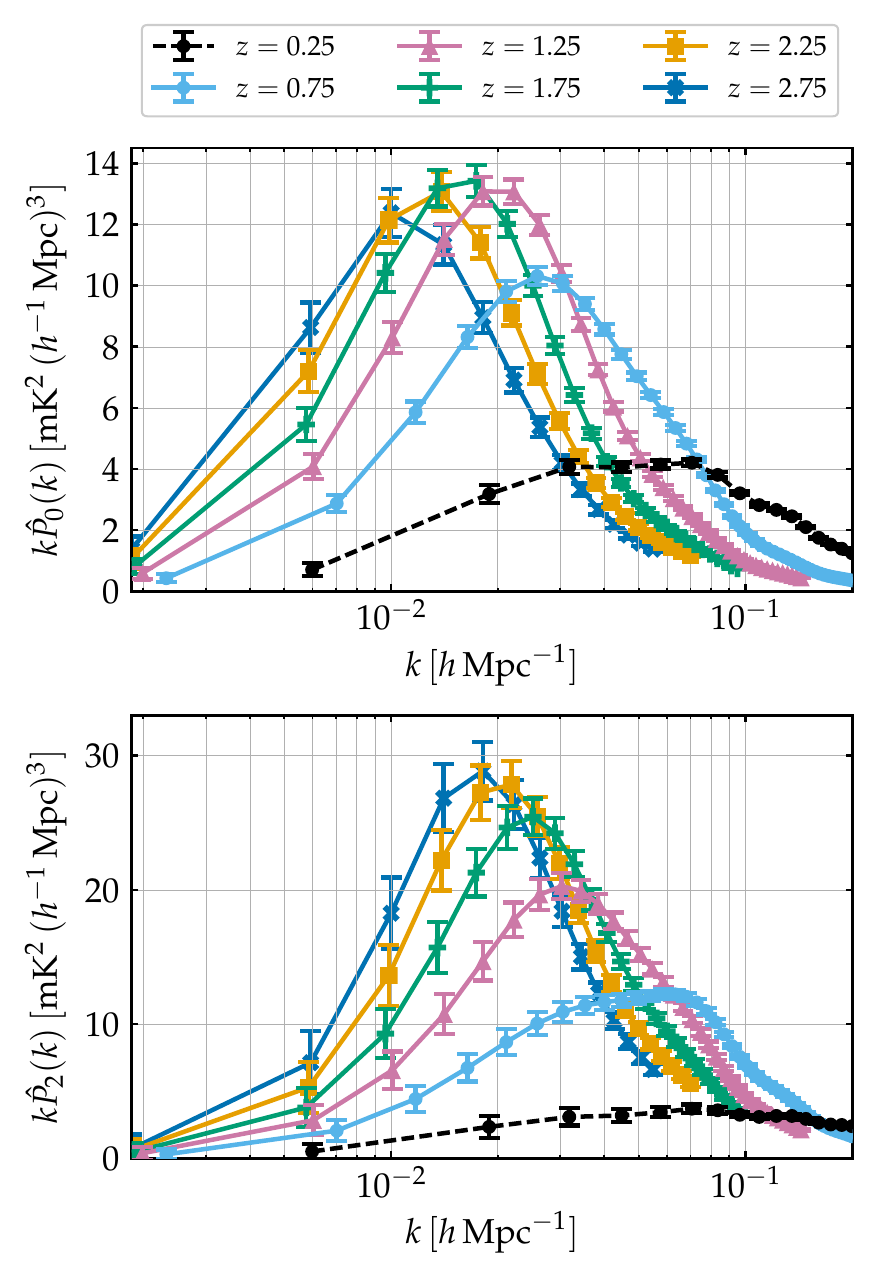}
    \caption{Tomographic mock data set for 21cm linear power spectrum monopole (upper panel) and quadrupole (lower panel) observations. The considered six redshift bins are centered at redshifts $\{0.25,\, 0.75,\, 1.25,\, 1.75,\, 2.25,\, 2.75 \}$. For the first redshift bin (black dashed line) we assume a SKA-Mid Band 2 survey. The data sets for the other five bins, instead, assume a SKA-Mid Band 1 survey (see \autoref{tab:SKA_specifics}). We refer to \autoref{sec:tomographic_data_set} for further details on how the signal and the errors are computed.}
    \label{fig:P_data_set}
\end{figure}
We follow \citet{Bacon:2018} to construct the mock data set and assume a combination of two surveys: a Medium-Deep Band 2 survey that covers a sky area of 5,000 deg$^2$ in the frequency range $0.95-1.75$ GHz (i.e. the redshift range $0-0.5$); a Wide Band 1 survey that covers a sky area of 20,000 deg$^2$ in the frequency range $0.35-1.05$ GHz (i.e. the redshift range $0.35-3$). 
We forecast observations for six equi-spaced, non-overlapping redshift bins, in the range $z=0-3$ with $\Delta z = 0.5$. The six bins are centered at redshifts $z_c = \{0.25,\, 0.75,\, 1.25,\, 1.75,\, 2.25,\, 2.75 \}.$ We assume the Band 2 survey specification for the mock 21cm power spectrum at redshift $0.25$ and Band 1 survey specification for all the others. Note that in our analysis each bin is regarded as independent.

\medskip
The survey sky coverage and the redshift range define the volume for the mock observations thus fixing the range of accessible scales for our analysis. 
In Fourier space, the largest scale available at each central redshift is $k_{\rm min}(z_c) = 2\pi / \sqrt[3]{V_{\rm bin}(z_c)}$, where $V_{\rm bin}(z_c)$ is the volume of each redshift bin, computed using \autoref{eq:volume_bin}. 

The smallest scale is, instead, imposed by the size of the telescope beam and it can be estimated as $k_{\rm max}(z_c) = 2\pi/R_{\rm beam}(z_c)$. At scales smaller than $k_{\rm max}$, the signal is dominated by the beam providing no relevant information on cosmology. Although we do not show results here, we tested pushing the $k_{\rm max}$ limit beyond the beam scale. We found no significant impact on the cosmological parameters constraints. To avoid entering the non-linear regime for the power spectrum we impose a cut-off at $k_{\rm max} = 0.2\, h \rm Mpc^{-1}$. 

Finally, we choose a fixed k-bin width as function of redshift $\Delta k(z_c)$ in order for to be large enough for modes to be independent, assuming $\Delta k(z_c)\sim 2k_{\rm min}(z_c)$.

\medskip
At each central redshift $z_c$ and data point $k$ we compute the monopole $\Hat{P}_0(z_c,k)$, the quadrupole $\Hat{P}_2(z_c,k)$ and the errors, as discussed in \secref{sec:mult_exp} and~\secref{sec:errors} respectively. In table~\autoref{tab:values_redshift}, we gather some of these redshift dependent quantities for the interested reader. The resulting forecasted data sets for the monopole and the quadrupole are shown in \autoref{fig:P_data_set}.

\begin{table*}
 \caption{Computed values of redshift dependent quantities at each central redshift $z_c$. For the first redshift bin ($z=0.25$) we assume SKA-Mid Band 2 specifications, while we use SKA-Mid Band 1 parameters for the other five bins. We refer to \autoref{tab:SKA_specifics} for a list of used SKA-Mid specifications.}
 \label{tab:values_redshift}
 \begin{tabular*}{2\columnwidth}{llccccccl} 
  \hline
   $z_{c}$ & central redshift & {0.25} & $0.75$ & $1.25$ & $1.75$ & $2.25$ & $2.75$\\
  \hline
    $\bar{T}_{\rm b} $  & mean brightness temperature & ${0.78\times 10^{-1}}$ & $1.3$ $\times 10^{-1}$& $1.9$ $\times 10^{-1}$& $2.5$ $\times 10^{-1}$& $3.0$ $\times 10^{-1}$ &  $3.6$ $\times 10^{-1}$ &[mK]\\
    $b_{\rm HI}$ & HI bias & {1.03}& $1.33$& $1.62$& $1.89$& $2.17$& $2.45$ \\
    $V_{\rm bin} $ & volume of the bin & $ {0.11\times 10^{10}}$ & 1.98 $ \times 10^{10}$& 3.14 $ \times 10^{10}$& 3.71 $ \times 10^{10}$&
       3.90 $ \times 10^{10}$ & 3.88 $ \times 10^{10}$ &$ [ h^{-3}\mathrm{Mpc}^3]$\\
    $R_{\rm beam}$ &  beam size  &  {6.38}&  23.36&  43.74&  66.13&
        89.83 & 114.5 & $ [h^{-1}\,\mathrm{Mpc}]$\\
    $P_{\rm N} $ & noise power spectrum  & {0.6} & 10 & 16 & 19 & 20 & 20 & $[\mathrm{mK^2}\, h^{-3}\mathrm{ Mpc^3 }]$\\
    $k_{\rm max}$  & minimum scale & {0.2}      & 0.2    & 0.14 & 0.09 & 0.07&
       0.05 &  $[h \rm Mpc^{-1}]$\\
    $k_{\rm min} $ & maximum scale& ${6.00\times 10^{-3}}$& 2.32$\times 10^{-3}$ & 1.99$\times 10^{-3}$& 1.88 $\times 10^{-3}$& 1.85 $\times 10^{-3}$& 1.85  $\times 10^{-3}$ & $[ h\, \rm Mpc^{-1}]$\\
    $\Delta k $ & $k$-bin width & ${12\times 10^{-3}}$ & 5 $\times 10^{-3}$& 4 $\times 10^{-3}$ & 4 $\times 10^{-3}$& 4 $\times 10^{-3}$& 4  $\times 10^{-3}$&$ [ h \rm Mpc^{-1}]$\\
    $N$ & number of data points & {16} & 43 & 36 & 25 & 18 & 14\\
  \hline
 \end{tabular*}
\end{table*}
\subsection{Constraining the cosmological parameters}
\label{sec:bayes}
\begin{figure}
	\includegraphics[width=\columnwidth]{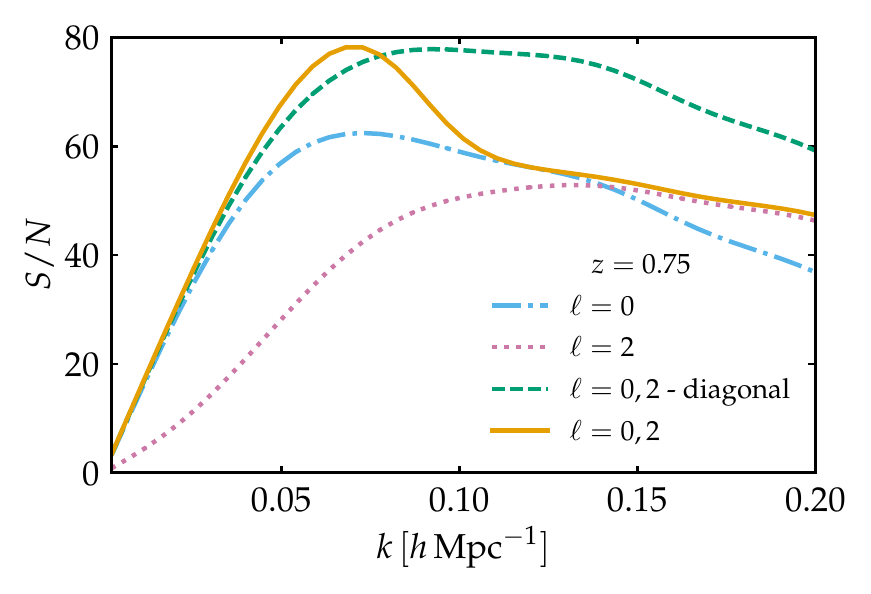}
		\includegraphics[width=\columnwidth]{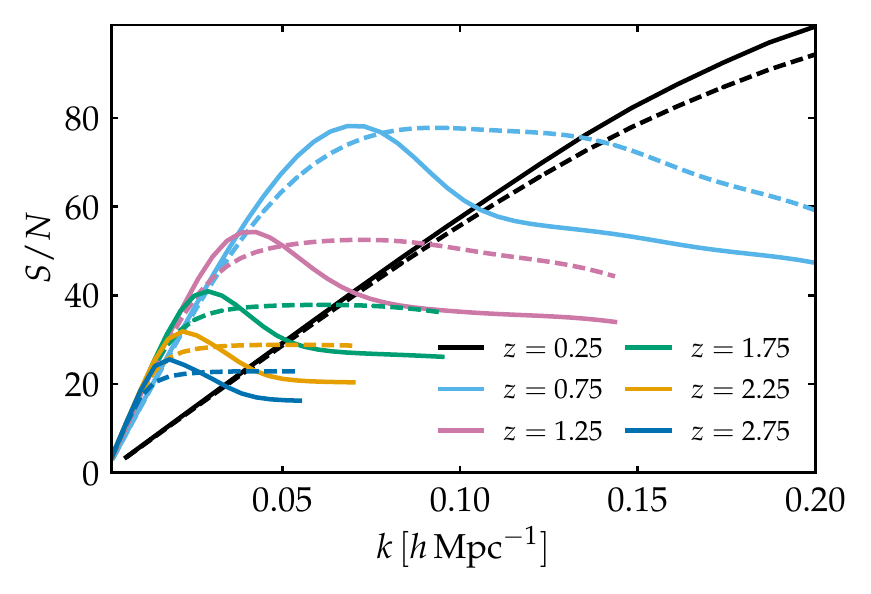}
    \caption{Signal-to-noise ratio as a function of $k$ (see \autoref{eq:signal_to_noise}). In the upper panel, we show the signal-to-noise at given redshift $z=0.75$, for different combinations of multipoles: the monopole alone (blue dashed-dotted line), the quadrupole alone (pink dotted line), the two combined (green dashed line) and the two combined considering the full non-diagonal covariance matrix (yellow solid line). In the lower panel, we show the signal-to-noise redshift dependence for the monopole and the quadrupole combined, considering a diagonal covariance matrix (dashed lines) and the full non-diagonal one (solid lines).}
    \label{fig:signal_to_noise}
\end{figure}

In order to exploit the constraining power of the mock data set presented in \secref{sec:tomographic_data_set}, we define a likelihood function (\secref{sec:likelihood}) and then set up the framework to constrain the cosmological parameters adopting a Bayesian approach (\secref{sec:numerical_analysis}). Given a set of observations and a theory that depends on set of parameters, the Bayes theorem links the posterior distribution to the likelihood function. The high-dimensional posterior can then be sampled using Monte-Carlo Markov-Chain (MCMC) methods~\citep[e.g.][]{gilks:1995,gamerman2006markov}.

\subsubsection{Likelihood function and signal-to-noise}
\label{sec:likelihood}
Given a set of measurements at scales $\{k_1, \dots, k_N\}$ and redshift $z_c$, to compute the likelihood function we define the vector 
\begin{equation}
    \bm{\Theta}(z_c) = \Big( \bm{\Hat{P}}_0(z_c), \bm{\Hat{P}}_2(z_c) \Big),
\end{equation}
with $\bm{\Hat{P}}_\ell(z_c) = (\Hat{P}_\ell(z_c,k_1), \dots, \Hat{P}_\ell(z_c,k_N))$.
When we use both the monopole and the quadrupole, the logarithmic likelihood is computed as
\begin{equation}
    \label{eq:likelihood}
    -\ln\big[ \mathcal{L}\big] = \sum_{z_c} \,\frac{1}{2} \,\Delta \bm{\Theta}(z_c)^{\rm T}\, \bm{\mathsf{C}}^{-1}(z_c)\,  \Delta \bm{\Theta}(z_c),
\end{equation}
where we define $\Delta \bm{\Theta}(z_c) = \bm{\Theta}^{\rm th}(z_c) - \bm{\Theta}^{\rm obs}(z_c)$, the difference between the values of $\bm{\Theta}(z_c)$ predicted from theory and observed. Here, $\bm{\mathsf{C}}(z_c)$ is the covariance matrix defined in \autoref{eq:full_covariance}, which becomes $\bm{\mathsf{C}}_{\rm diag}(z_c)$, i.e. \autoref{eq:covariance_diagonal}, when we neglect multipole covariance. We consider independent redshift bins, i.e. we simply sum over the contribution from each central redshift. When studying $\Hat{P}_0$ and $\Hat{P}_2$ separately we use only the relevant blocks of the covariance matrix $\bm{\mathsf{C}}(z_c)$, thus using a simplified version of \autoref{eq:likelihood}.

\medskip
Using a similar formalism, we can compute the signal-to-noise at a specific $k$ for each central redshift as
\begin{equation}
\label{eq:signal_to_noise}
    [\mathrm{S/N}]^2(z_c,k) = \bm{\Theta}_k(z_c)^{\rm T}\, \bm{\mathsf{C}}_k^{-1}(z_c)\, \bm{\Theta}_k(z_c)
\end{equation}
where $\bm{\Theta}_k(z_c) = (\Hat{P}_0(z_c,k), \Hat{P}_2(z_c,k))$ and $\bm{\mathsf{C}}_k(z_c)$ is a matrix defined as 
\begin{equation}
 \bm{\mathsf{C}}_k(z_c) =   \begin{pmatrix}
    C_{00}(z_c,k) & C_{02}(z_c,k) \\
     C_{02}(z_c,k) & C_{22}(z_c,k)
    \end{pmatrix}.
\end{equation}
The expression of the signal-to-noise when we neglect the multipole covariance or when we use only $\Hat{P}_0$ or $\Hat{P}_2$ is modified accordingly, as described for the likelihood function above. 

\medskip
The resulting signal-to-noise as a function of $k$ is shown in \autoref{fig:signal_to_noise}. We recall that that the maximum scales explored is the minimum scale between the maximum scale imposed by the beam width and the linear regime cut-off of $k_{\rm max} = 0.2\, h \rm Mpc^{-1}$ (see \secref{sec:tomographic_data_set}).

At fixed redshift (upper panel of \autoref{fig:signal_to_noise}), we observe that, when the monopole and the quadrupole are used together ($\ell =0,2$), we get a higher signal-to-noise with respect to the monopole ($\ell =0$) and the quadrupole ($\ell=2$) alone. For the $\ell =0,2$ case, we observe that when we consider multipole covariance (yellow solid line) we get an enhancement of the signal-to-noise at higher scales and a suppression at lower ones. As discussed in~\cite{Soares:2020}, this effect is caused by the beam smoothing factor in the model for $\Hat{P}_{21}$ (see \secref{sec:beam_model}). 
We examine the implications of using different combinations of multipoles on the parameters constraints in \secref{sec:res_P21_P21+Planck}. 

We show how the signal-to-noise decreases as a function of redshift in the lower panel of \autoref{fig:signal_to_noise}. Its shape is consistent at all redshifts (the full signal-to-noise for the mock measurements in the Band 2 ($z=0.25$), can be found in~\autoref{fig:non_linear}). In our analysis we will always consider the cumulative contribution from all the redshift bins.

\subsubsection{Numerical analysis}
\label{sec:numerical_analysis}
To perform the MCMC analysis we use an expanded version of the MCMC sampler \texttt{CosmoMC}\footnote{See \url{https://cosmologist.info/cosmomc}.}~\citep{Lewis:2002,Lewis:2013}. We modify it in order to include the computation of the theoretical expectations for the monopole and the quadrupole (see \appref{sec:app_equations}) and of the likelihood function defined above (see \autoref{eq:likelihood}). 

The analysis of the MCMC samples to compute the marginalised constraints is performed with the Python package \texttt{GetDist}\footnote{See \url{https://getdist.readthedocs.io}.}~\citep{Lewis:2019xzd}.

We conduct an MCMC analysis varying the six parameters describing the $\Lambda$CDM model, i.e. we vary $\{\Omega_b h^2,\, \Omega_c h^2, n_s,\, \ln (10^{10} A_s),\, \tau,\, 100\theta_{\rm MC} \}$. Results on other parameters, such as $H_0$ and $\sigma_8$, are derived from results on this set. The assumed fiducial cosmology is Planck 2018~\citep{planck:2018}. We list the fiducial values and the flat prior we use in \autoref{tab:fiducial_cosmology}.

\subsubsection{CMB data sets}
\label{sec:Planck_data_set}

In this study, we combine our mock 21cm IM data set with Planck 2018~\citep{planck:2018}. The CMB likelihood includes the high-$\ell$ TT, TE, EE lite likelihood in the interval of multipoles $30\leq\ell\leq 2508$ for TT and $30\leq\ell\leq 19696$ for TE, EE. Lite likelihoods are  calculated with the \texttt{Plik lite} likelihood~\citep{planck:2018like}. Instead for the low-$\ell$ TT power spectrum we use data from the \texttt{Commander} component-separation algorithm in the range $2 \leq \ell \leq 29$. We adopt also the Planck CMB lensing likelihood and the low EE polarization power spectrum, referred to as lowE, in the range $2 \leq \ell\leq 29$, calculated from the likelihood code \texttt{SimAll}~\citep{planck:2018_maps}. In the rest of the paper with the label "Planck 2018" we refer to the combination TT, TE, EE + low-$\ell$ + lowE + lensing.

\section{Results}
\label{sec:results}
\begin{table}
	\centering
	\caption{Assumed fiducial cosmology~\citep{planck:2018} and used flat priors.}
	\label{tab:fiducial_cosmology}
	\begin{tabular*}{\columnwidth}{l@{\hspace*{0.8in}}c@{\hspace*{0.7in}}c@{\hspace*{0.7in}}} 
	\hline 
	Parameter & Fiducial value & Prior\\
		\hline
        $\Omega_b h^2$ & 0.022383 & $[0.005, 0.1]$\\
        $\Omega_c h^2$ & 0.12011 & $[0.001, 0.99]$\\
        $n_s$ & 0.96605 & $[0.8, 1.2]$\\
        $\ln (10^{10} A_s)$ & 3.0448 & $[1.61, 3.91]$\\
        $\tau$ & 0.0543 & $[0.01, 0.8]$\\
        $100\theta_{\rm MC}$ & 1.040909 & $[0.5, 10]$\\
		\hline
	\end{tabular*}
\end{table}
We present in this section the results of our analysis. We first explore the constraining power of the mock 21cm data set, using different combinations of multipoles (\secref{sec:res_P21}); we then combine the mock data set with Planck CMB data (\secref{sec:res_P21_P21+Planck}); we study
the effect of nuisance parameters describing the neutral hydrogen astrophysics in \secref{sec:res:nuisances}; finally, we discuss the impact of non-linear scales in \secref{sec:res_non_linear}.

We show the marginalized 1D and 2D posteriors for the studied set of parameters. Note that 68\% confidence level constraints are presented as percentages with respect to the marginalized mean value. 

\subsection{Probing the constraining power of 21cm signal observations}
\label{sec:res_P21}
\begin{figure}
	\includegraphics[width=\columnwidth]{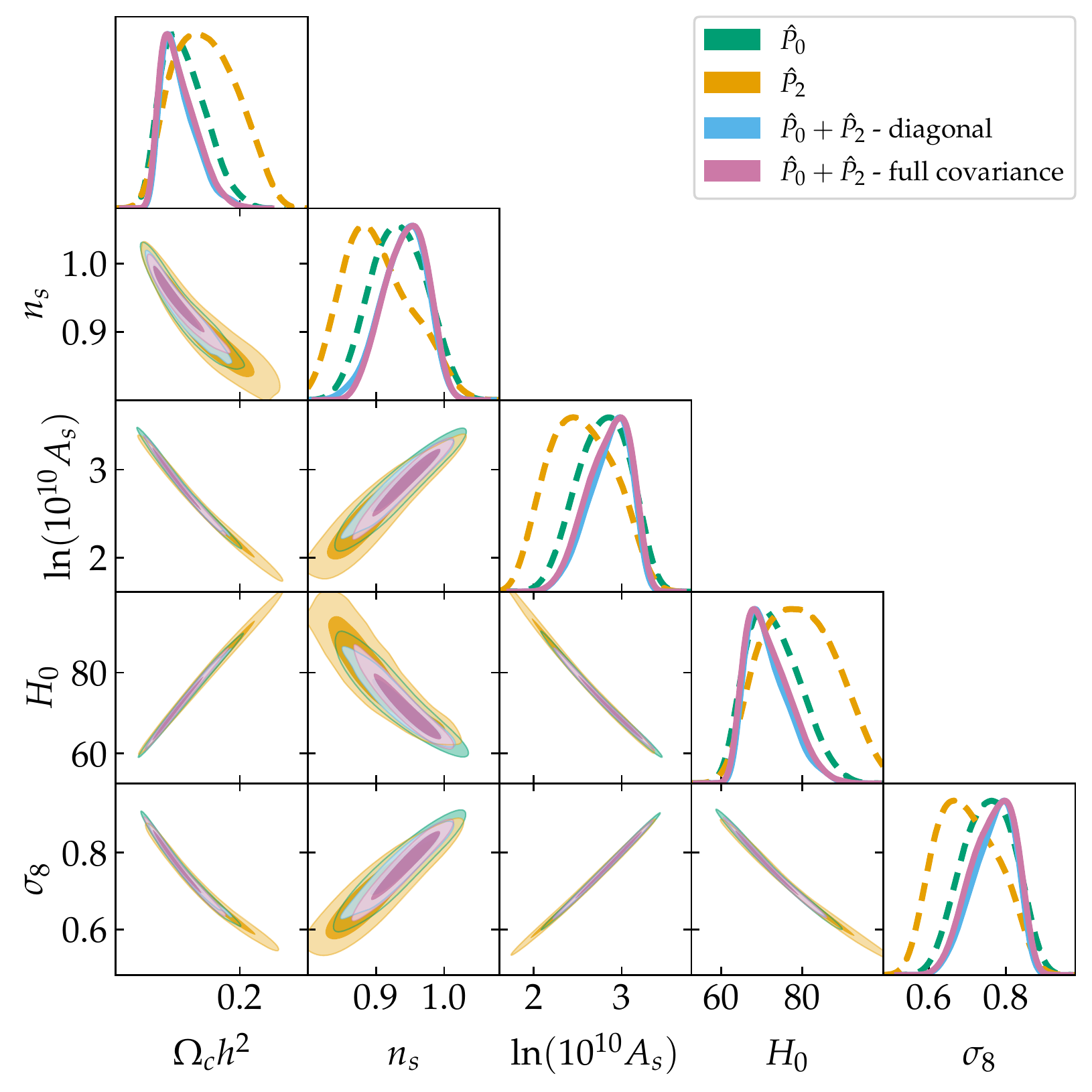}
    \caption{Joint constraints (68\% and 95\% confidence regions) and marginalized posterior distributions on a subset of the cosmological parameters. We show the constraints obtained using the mock tomographic data set for the monopole only ("$\Hat{P}_0$"), the quadrupole only ("$\Hat{P}_2$"), the two combined with ("$\Hat{P}_0 + \Hat{P}_2$ non diagonal") and without ("$\Hat{P}_0 + \Hat{P}_2$ diagonal") considering multipole covariance. The relative constraints are listed in \autoref{tab:constraints_P21_alone}. }
    \label{fig:results_P21}
\end{figure}

In \autoref{fig:results_P21} and \autoref{tab:constraints_P21_alone}, we show the marginalized contours and constraints\footnote{We specify that, when dealing with asymmetrical posterior distributions, we estimate the percentage constraint using the mean value between the left and right marginalized error. Given that this is a forecasts analysis, this approximation is enough for the purpose of this work.} that we obtain using our SKA-Mid tomographic data set alone, for different combinations of multipoles. Note that we show only some of the model parameters for brevity.

\begin{figure*}
	\includegraphics[width=2\columnwidth]{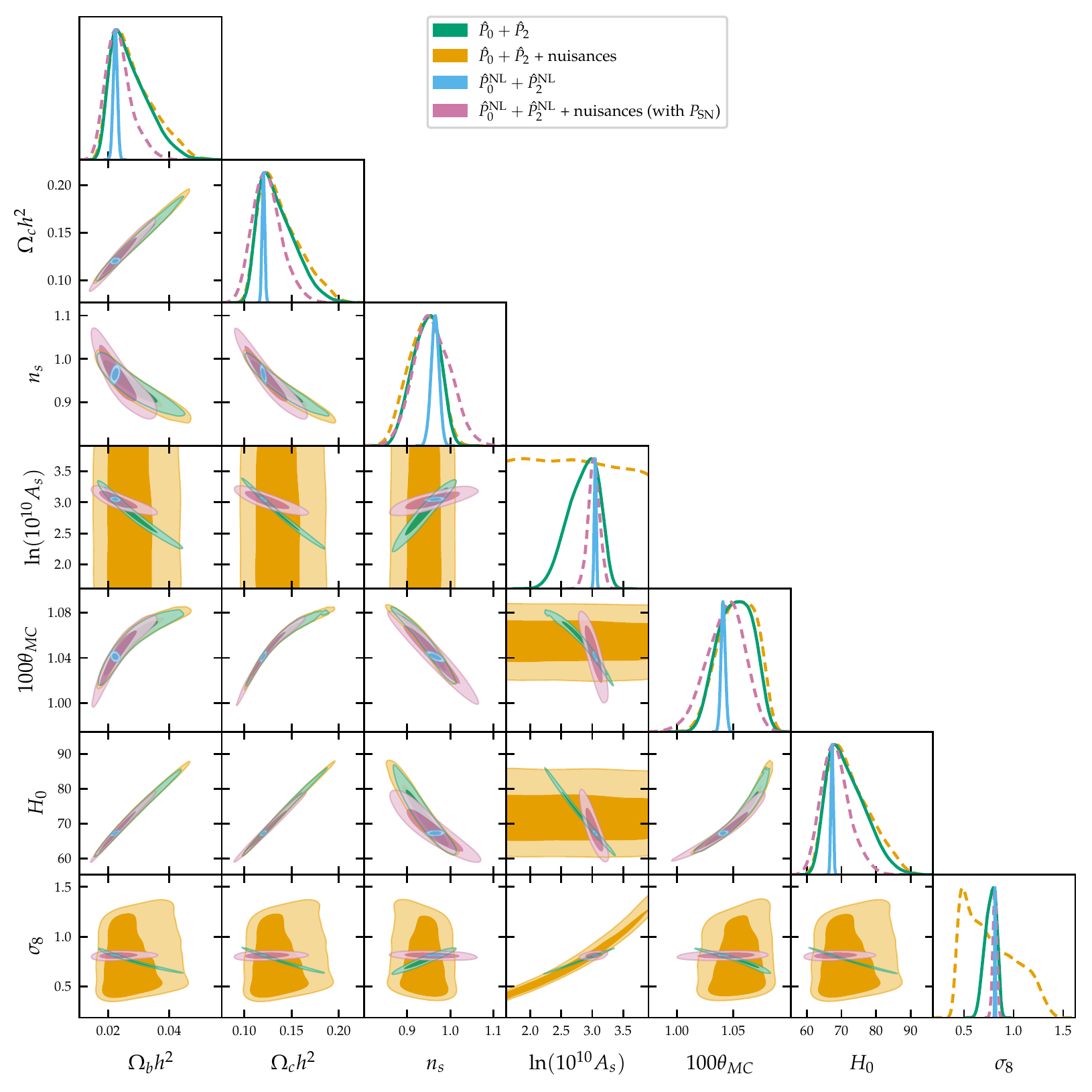}
    \caption{Joint constraints (68\% and 95\% confidence regions) and marginalized posterior distributions on cosmological parameters. Here, the label "$\Hat{P}_0 + \Hat{P}_2$" stands for the baseline tomographic data set for the monopole and the quadrupole combined and with multipole covariance taken into account. The label "$\Hat{P}_0^{\rm NL} + \Hat{P}_2^{\rm NL}$" indicates that the full non-linear data set has been used. The label "nuisances" (dashed lines) indicates that we vary the nuisances parameters along with the cosmological ones. The relative constraints are listed in \autoref{tab:constraints_P21}.}
    \label{fig:full_P21}
\end{figure*}

As one could expect from the signal-to-noise prediction of  \autoref{fig:signal_to_noise}, using only the quadrupole leads to the broadest constraints, while the most constraining results are obtained for the monopole and the quadrupole combined. 
The off-diagonal terms of the multiple covariance do not affect much the constraints. The marginalized percentage constraints for the baseline case ($\Hat{P}_0 + \Hat{P}_2$ considering the full covariance) for the complete set of cosmological parameters can be found in \autoref{fig:full_P21} and \autoref{tab:constraints_P21}.

\begin{table}
	\centering
	\caption{Marginalized percentage constraints on cosmological parameters at the 68\% confidence level. We show the results obtained using the mock tomographic data set for the monopole only ("$\Hat{P}_0$"), the quadrupole only ("$\Hat{P}_2$"), the two combined with ("$\Hat{P}_0 + \Hat{P}_2$ full covariance") and without ("$\Hat{P}_0 + \Hat{P}_2$ diagonal") considering multipole covariance. Confidence regions for the same set of results are shown in \autoref{fig:results_P21}.}
	\label{tab:constraints_P21_alone}
	\begin{tabular}{lcccc} 
	\hline 
	\multirow{2}{*}{Parameter} & \multirow{2}{*}{$\Hat{P}_0$} & \multirow{2}{*}{$\Hat{P}_0$}& $\Hat{P}_2 + \Hat{P}_2$ & $\Hat{P}_0 + \Hat{P}_2$ \\
    	& & & diagonal & full covariance \\
		\hline
$\Omega_c h^2$ & 16.71\%& 21.57\%& 12.71\%& 13.36\% \\
$n_s$ & 4.59\%& 5.59\%& 3.55\%& 3.44\% \\
${\rm{ln}}(10^{10} A_s)$ & 10.94\%& 15.26\%& 8.26\%& 8.83\% \\
$H_0$ & 9.09\%& 12.01\%& 6.91\%& 7.39\% \\
$\sigma_8$ & 9.56\%& 11.92\%& 7.11\%& 7.64\% \\
		\hline
	\end{tabular}
\end{table}

Our results show, as expected, that the constraining power on the cosmological parameters of the SKA-Mid mock data set is greater than what one could obtain with MeerKAT alone~\citep{Berti:2021}. The data set we constructed is enough to constrain five out of six of the cosmological parameters. This is because 21cm observations are not sensitive to variations on $\tau$, that remains unconstrained. 
The marginalised confidence levels are broad with respect to Planck constraints~\citep{planck:2018}, but comparable with other probes. E.g., with tomographic observations of the monopole and the quadrupole combined we constrain $H_0$ with a relative error of $\sigma_{H_0} =7.4\%$,
\begin{equation}
\label{eq:H0_linear}
   H_0 =  71.6^{+3.8}_{-6.8}\,\, \text{ km s}^{-1}\text{Mpc}^{-1} \quad {\scriptstyle(68\%,\ \Hat{P}_0 + \Hat{P}_2 \text{ - full covariance})}.
\end{equation}
We stress that when we state a constraint on a single parameter obtained using our mock 21cm data set, the central value does not have any physical meaning and it is driven by the input fiducial cosmology value.

Although not competitive with Planck, our tomographic measurements with six redshift bins and at linear scales provide an estimate of $H_0$ with an uncertainty comparable with others late Universe measurements~\citep{Verde:2019}, and, as we further discuss in \secref{sec:res_non_linear}, constraints on $H_0$ are improved if we extend our data set to non-linear scales. SKA-Mid 21cm observations will have thus the potential to provide new information for the discussion on the $H_0$ value~\citep{Schoneberg:2021qvd}.

Looking at the 2D contours, we observe that there is a marked degeneracy between the cosmological parameters. As already found for mock MeerKAT observations, 21cm measurements show a strong degeneracy in the $H_0$ - $\Omega_ch^2$ plane~\citep{Berti:2021}. This feature is pivotal when combining intensity mapping data with CMB measurements, as we discuss in the following section.

\subsection{21cm signal observations combined with CMB data}
\label{sec:res_P21_P21+Planck}
In the above section we studied the constraining power of the 21cm multipoles. Here, we combine our baseline data set (the monopole $\Hat{P}_0$ plus the quadrupole $\Hat{P}_2$ considering the full covariance) with Planck CMB measurements. The rationale behind this is to investigate if and how 21cm observations can complement the detailed information on the cosmological parameters carried by the CMB. 

\medskip
We refer to \secref{sec:Planck_data_set} for a description of the used Planck 2018~\citep{planck:2018} data sets and likelihoods. For consistency, 
we first run the Planck likelihood in our framework and reproduce constraints in agreement with the Planck 2018 results
\begin{equation}
    \begin{rcases}
    \Omega_c h^2 = 0.1201\pm 0.0012 \\
   H_0 = 67.32 \pm 0.53\ \text{ km s}^{-1}\text{Mpc}^{-1}\\
      {\rm{ln}}(10^{10} A_s) = 3.045\pm 0.014 \\
   \sigma_8 = 0.8115\pm 0.0060
    \end{rcases}
    \ {\scriptstyle(68\%,\ \text{Planck 2018}).}
\end{equation}

\begin{figure*}
	\includegraphics[width=2\columnwidth]{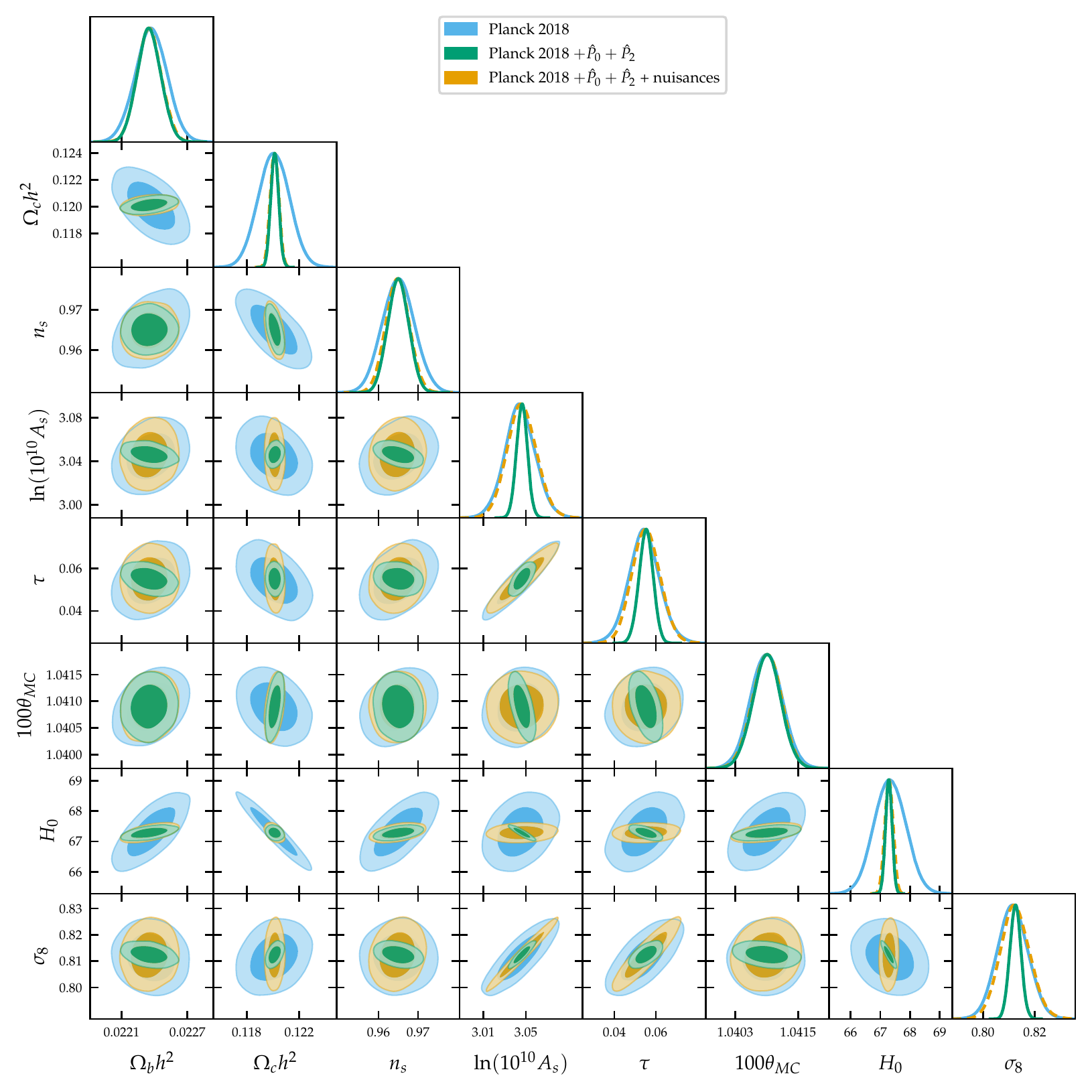}
    \caption{Joint constraints (68\% and 95\% confidence regions) and marginalized posterior distributions on cosmological parameters. The label "Planck 2018" stands for TT, TE, EE + lowE + lensing, while the label "$\Hat{P}_0 + \Hat{P}_2$" stands for the baseline tomographic data set for the monopole and the quadrupole combined and with multipole covariance taken into account. The label "nuisances" (dashed line) indicates that we vary the nuisances parameters along with the cosmological ones. The relative constraints are listed in \autoref{tab:constraints_P21_Planck}.}
    \label{fig:full_results}
\end{figure*}

\medskip
Our results for the combination of the CMB data and our mock 21cm observation are presented in \autoref{tab:constraints_P21_Planck} and in \autoref{fig:full_results}.
Adding the 21cm power spectrum multipoles to the CMB, significantly improves the constraining power on the majority of the cosmological parameters. The effect is particularly pronounced on $\Omega_c h^2$ and $H_0$, for which the error is reduced by approximately a fourth. This gain in constraining power is due to the combination of opposite degeneracy directions between the CMB and the 21cm power spectrum on these cosmological parameters. 
This effect is particularly strong in the $H_0$ - $\Omega_ch^2$ plane, where the degeneracy is completely removed. In $A_s$ - $\sigma_8$ plane, the effect is milder, but still significant. 

\medskip
In more detail, we find
\begin{equation}
    \begin{rcases}
    \Omega_c h^2 = 0.12014\pm 0.00030 \\
   H_0 = 67.28\pm 0.11\ \text{ km s}^{-1}\text{Mpc}^{-1}\\
       {\rm{ln}}(10^{10} A_s) = 3.0463\pm 0.0052 \\
   \sigma_8 = 0.8125\pm 0.0021
    \end{rcases}
    \ {\scriptstyle(68\%,\ \text{Planck 2018} + \Hat{P}_0 + \Hat{P}_2).}
\end{equation}
We recall that the central value of the obtained constraints does not have a physical meaning and it is driven by the input fiducial cosmology we use for our mock 21cm observations. However, these values are useful to properly visualize the constraining power of our mock observations. 

More important from a quantitative point of view is, instead, the relative error. 
We find $\sigma_{\Omega_ch^2} = 0.25\%$ and $\sigma_{H_0} = 0.16\%$, to be compared with the Planck only estimates of $\sigma_{\Omega_ch^2} = 0.99\%$ and $\sigma_{H_0} = 0.79\%$. 
The estimate of the error on $H_0$ we obtain combining 21cm power spectrum multipoles with CMB is competitive with other LSS probes, e.g. with Euclid\footnote{\url{https://sci.esa.int/web/euclid}} forecasts~\citep{Euclid:2019clj}. The errors on $A_s$ and $\sigma_8$ are significantly reduced too, by more than a factor two: the relative errors are $\sigma_{\ln (10^{10} A_s)} = 0.17\%$ and $\sigma_{\sigma_8} = 0.26\%$ to be compared with $\sigma_{\ln (10^{10} A_s)} = 0.46\%$ and $\sigma_{\sigma_8} = 0.73\%$ of the Planck only result.
Moreover, it is interesting to see how the improvement on the other cosmological  parameters induces a better estimate of $\tau$, although the 21cm observable alone has not a significant constraining power on it. 

We conclude that 21cm observations provide complementary information to the CMB, allowing for a significantly improved estimation of the cosmological parameters. 

Note that the improvement is stronger than the effect of adding BAO measurements to the CMB~\citep{planck:2018}. Although we do not show results here, we tested also the effect of using BAO~\citep{Beutler:2011,Ross:2014,BOSS:2016} along with the multipoles and the CMB, finding no significant repercussion on the constraints.

\begin{figure}
	\includegraphics[width=1\columnwidth]{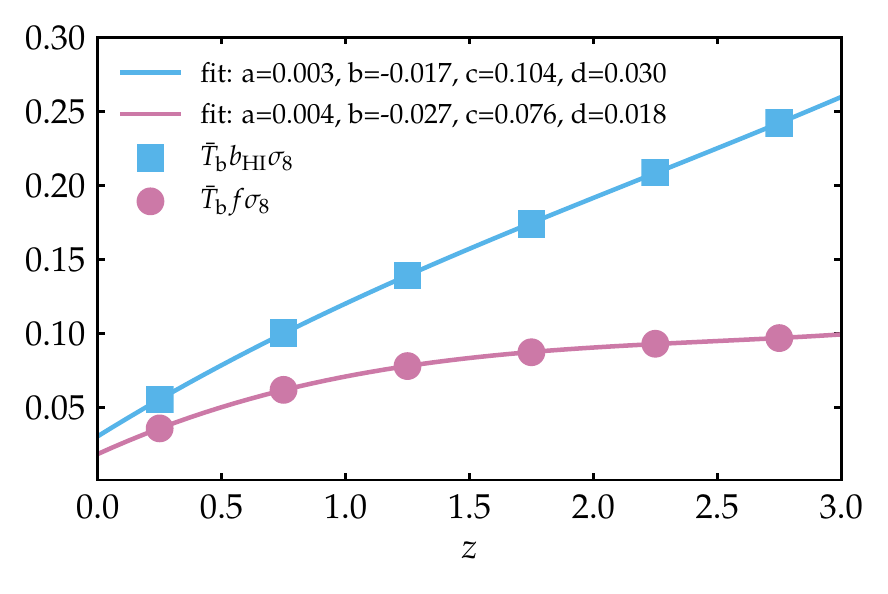}
    \caption{Redshift evolution for the nuisance parameters $\bar{T_{\rm b}} b_{\rm HI} \sigma_8(z) $ and $\bar{T_{\rm b}} f \sigma_8 (z)$. We show the theory predicted values for the six redshift bins we consider (circles and squares) and the best-fit redshift evolution (solid lines) modelled as a $3^{\rm rd}$-degree polynomial (see \autoref{eq:fitting_nuisances}).}
    \label{fig:nuisances_redshift}
\end{figure}

\begin{table}
	\centering
	\caption{Marginalized percentage constraints on cosmological parameters at the 68\% confidence level. Here, the label "Planck 2018" stands for TT, TE, EE + lowE + lensing, while the label "$\Hat{P}_0 + \Hat{P}_2$" stands for the baseline tomographic data set for the monopole and the quadrupole combined and with multipole covariance taken into account. The label "nuisances" indicates that we vary the nuisances parameters along with the cosmological ones. Confidence regions for the same set of results are shown in \autoref{fig:full_results}.}
	\label{tab:constraints_P21_Planck}
	\begin{tabular}{lccc} 
	\hline 
    Parameter & Planck 2018 & $+ \Hat{P}_0 + \Hat{P}_2$& $+$ nuisances \\
		\hline
$\Omega_b h^2$ & 0.64\%& 0.49\%& 0.49\% \\
$\Omega_c h^2$ & 0.99\%& 0.25\%& 0.27\% \\
$n_s$ & 0.42\%& 0.27\%& 0.31\% \\
${\rm{ln}}(10^{10} A_s)$ & 0.46\%& 0.17\%& 0.45\% \\
$\tau$ & 13.44\%& 6.09\%& 12.19\% \\
$100\theta_{MC}$ & 0.03\%& 0.03\%& 0.03\% \\
$H_0$ & 0.79\%& 0.16\%& 0.20\% \\
$\sigma_8$ & 0.73\%& 0.26\%& 0.70\% \\
		\hline
	\end{tabular}
\end{table}

\subsection{Introducing astrophysical uncertainties}
\label{sec:res:nuisances}
In the analysis discussed above, we assumed a perfect knowledge of the astrophysics involved in the estimate of 21cm signal observations. In particular, we assumed to know the total HI density $\Omega_{\rm HI}$ (that enters in \autoref{eq:Tb}) and the HI bias $b_{\rm HI}$ as a function of redshift. However, these quantities depend on the detailed baryon physics at play and their connection with dark matter is not completely understood \citep[e.g.][]{Zoldan2017,Guo2017,Villaescusa-Navarro:2018,Spinelli:2019}. 
To take into account our ignorance on these parameters in our analysis, we follow \citet{Bernal:2019} and rewrite the power spectrum of \autoref{eq:P_21_full} as
\begin{align}
    \begin{split}
        \Hat{P}_{21}(z,\,k,\, \mu) &= \tilde{B}^2 (z,k,\mu) \Big[ \Bar{T}_{\rm b}(z)b_{\mathrm{HI}}(z)\sigma_8(z) + \\
        & \qquad +\Bar{T}_{\rm b}(z)f(z)\sigma_8(z)\, \mu^2\Big]^2 \frac{P_{\rm m}(z,k)}{\sigma_8(z)}.
    \end{split}
\end{align}
The redshift dependent combinations of functions $\Bar{T}_{\rm b}b_{\mathrm{HI}}\sigma_8(z)$ and $\Bar{T}_{\rm b}f\sigma_8(z)$ can be added to the set of estimated parameters as nuisances. 

The most general parametrization for these nuisance parameters does not impose any specific redshift evolution. Given that we have six redshift bins, we need twelve new parameters: six $[\Bar{T}_{\rm b}b_{\mathrm{HI}}\sigma_8]_i$ and six $[\Bar{T}_{\rm b}f\sigma_8]_i$, one for each redshift, with $i$ being $i=\{1,\dots,6\}$. However, the high dimensionality of this configuration impact significantly the required computational time for the convergence of the MCMC procedure for the exploration of the posterior.

Alternatively, one can lower the number of nuisance parameters by assuming a parametrization for the redshift evolution of $\Bar{T}_{\rm b}b_{\mathrm{HI}}\sigma_8(z)$ and $\Bar{T}_{\rm b}f\sigma_8(z)$ in agreement with their theoretical prediction. We use a $3^{\rm rd}$-degree polynomial model 
\begin{equation}
    \label{eq:fitting_nuisances}
    \Bar{T}_{\rm b}b_{\mathrm{HI}}\sigma_8(z), \: \Bar{T}_{\rm b}f\sigma_8(z) = az^3 + bz^2 + cz + d,
\end{equation}
and reduce the nuisances from twelve to eight:  four coefficients $[\Bar{T}_{\rm b}f\sigma_8]_q$, with $q=\{a,b,c,d\}$, and other four $[\Bar{T}_{\rm b}f\sigma_8]_q$. We find that assuming this redshift evolution gives the same results with respect to the twelve nuisances case while we achieve a better and faster convergence. Thus, we choose to work with this latter parametrization of the nuisances.

In \autoref{fig:nuisances_redshift} we show the theoretical redshift evolution (under the assumption discussed in \secref{sec:P21_model}) and the fitted one for $\Bar{T}_{\rm b}b_{\mathrm{HI}}\sigma_8$ and $\Bar{T}_{\rm b}f\sigma_8$. In the following we show the results we obtain varying the eight nuisances parameters.
Note that we assume a very wide flat prior, centered at the theoretical expected value for each nuisance parameter.

We present the results for the multipoles alone in \autoref{tab:constraints_P21} and \autoref{fig:full_P21}. 

\medskip
When openning the parameter space to the nuisances, we see a complete loss of constraining power on $A_s$. This is expected, due to the fact that varying the nuisances we loose information on the amplitude of the power spectrum. The deterioration of the constraint on $A_s$ 
translates in a weakening of the constraining power on $\sigma_8$. 

Nevertheless, the impact of nuisance parameters is limited to these two parameters. The constraints on the other cosmological parameters remain unaffected showing the power of tomography: using the six redshift bins allows to include the evolution of the 21cm power spectrum multipoles and, thus, preserves their constraining power in particular on $\Omega_ch^2$ and $H_0$.

\medskip
The same discussion applies when we combine the multipoles with CMB, as in \secref{sec:res_P21_P21+Planck} but also varying the nuisance parameters. Results are shown in \autoref{fig:full_results} and \autoref{tab:constraints_P21_Planck} from which it can be seen that the constraints on $\Omega_ch^2$ and $H_0$ remain essentially unvaried. Note, however, that the constrain on $A_s$ and  $\tau$ and, consequently, on $\sigma_8$ are driven just by the Planck data.

For completeness, 2D contours and the marginalized posteriors for the nuisance parameters themselves are shown in \autoref{fig:full_nuisances} and discussed in \appref{sec:app_nuisances}.

\subsection{Extending to non-linear scales}
\label{sec:res_non_linear}
Up to now, we investigated the constraining power on the cosmological parameters of 21cm observations at linear scales that are the once best sampled by the large beam of the single-dish intensity mapping. At the linear scales, it is also possible to explore beyond $\Lambda$CDM models \citep{Berti:2021}, for which we often lack non-linear scale predictions. In a $\Lambda$CDM scenario and for the low-z bins, we can, however, push our analysis to larger $k$ and study their constraining power. 

For all redshift bins but the first two, the $k_{\rm max}$ cut-off due to the frequency dependent beam (see \secref{sec:tomographic_data_set}) is much stronger than the linear-scale cut-off $k = 0.2 \, h \, \text{Mpc}^{-1}$. The two lowest redshift bins ($z=0.25$ and $z=0.75$) can instead be extended to larger $k$ if we relax the linear-scale cut off. We acquire 15 and 67 new points and we are able to reach $k \sim 0.27 \, h \, \text{Mpc}^{-1}$ and 
$k \sim 1 \, h \, \text{Mpc}^{-1}$ at redshifts $z=0.75$ and $z=0.25$, respectively.
In this new k-range the shot noise is non-negligible and it needs to be considered in the modelling.

\medskip
We create the new mock non-linear data set as
\begin{align}
\begin{split}
    \label{eq:P_21_full_NL}
     \Hat{P}_{21}^{\rm NL}(z,\,k,\, \mu) &= \tilde{B}^2 (z,k,\mu) \big[ {P}_{21}^{\rm NL}(z,k) + {P}_{\rm SN}(z)\big],
\end{split}
\end{align}
where ${P}_{\rm SN}$ is the shot noise level estimated at different redshift interpolating results from hydro-dynamical simulations~\citep{Villaescusa-Navarro:2018}.
The non-linear 21cm power spectrum $P_{21}^{\rm NL}(z,k)$ is obtained as in~\autoref{eq:P21}, but substituting the linear matter power spectrum with the non-linear one, computed numerically with \texttt{CAMB}\footnote{We use the \texttt{HALOFIT}~\citep{Smith:2002} version from~\cite{Mead:2016}.}. The expressions for the 21cm multipoles are changed accordingly. 
\begin{table}
	\centering
	\caption{Marginalized percentage constraints on cosmological parameters at the 68\% confidence level. Here, the label "$\Hat{P}_0 + \Hat{P}_2$" stands for the baseline tomographic data set for the monopole and the quadrupole combined and with multipole covariance taken into account. The label "$\Hat{P}_0^{\rm NL} + \Hat{P}_2^{\rm NL}$" indicates that the full non-linear data set has been used. The label "nuis." indicates that we vary the nuisances parameters along with the cosmological ones. The symbol "---" stands for unconstrained. Confidence regions for the same set of results are shown in \autoref{fig:full_P21}.}
	\label{tab:constraints_P21}
	\begin{tabular}{lcccc} 
	\hline 
    Parameter & $\Hat{P}_0 + \Hat{P}_2$ & $+$ nuis. & $ \Hat{P}_0^{\rm NL} + \Hat{P}_2^{\rm NL}$ & $+$ nuis. ($\Hat{P}_{\rm SN}$) \\
		\hline
$\Omega_b h^2$ & 21.04\%& 22.81\%& 3.02\%& 17.30\% \\
$\Omega_c h^2$ & 13.36\%& 14.66\%& 1.16\%& 12.27\% \\
$n_s$ & 3.44\%& 3.94\%& 0.95\%& 4.45\% \\
${\rm{ln}}(10^{10} A_s)$ & 8.83\%& --- & 0.49\%& 3.00\% \\
$100\theta_{MC}$ & 1.53\%& 1.62\%& 0.18\%& 1.61\% \\
$H_0$ & 7.39\%& 8.10\%& 0.49\%& 5.90\% \\
$\sigma_8$ & 7.64\%& --- & 0.37\%& 2.41\% \\

		\hline
	\end{tabular}
\end{table}

\begin{figure}
	\includegraphics[width=1\columnwidth]{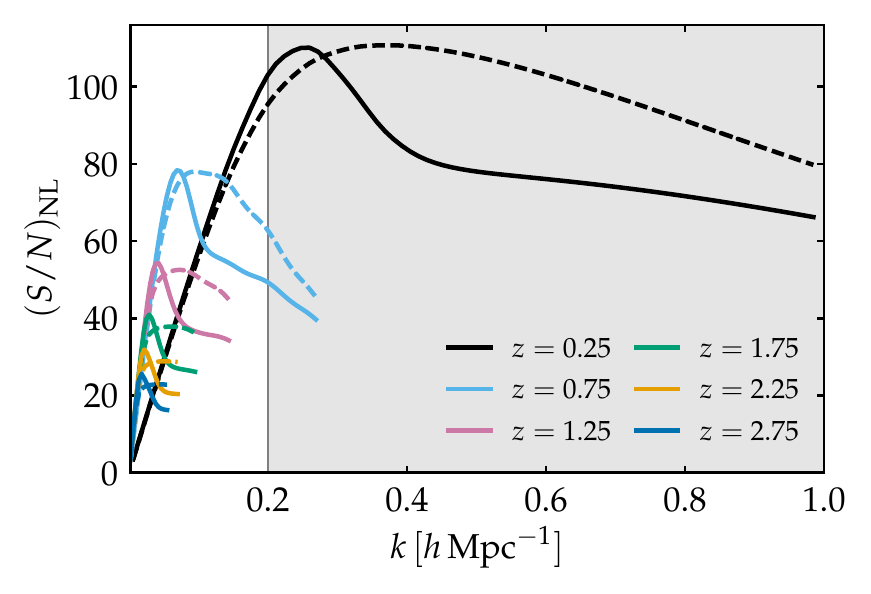}
    \caption{Computed signal-to-noise ratio as a function of $k$ (see \autoref{eq:signal_to_noise}). We show the signal-to-noise computed for the six redshift bins and for the monopole and the quadrupole combined, considering a diagonal covariance matrix (dashed lines) or the full non-diagonal one (solid lines). The shaded area highlights the new scales acquired extending the mock 21cm power spectrum to non-linear scales.}
    \label{fig:non_linear}
\end{figure}

\medskip
In \autoref{fig:non_linear}, we show the signal-to-noise for the new non-linear tomographic data set obtained with the model of~\autoref{eq:P_21_full_NL}. 
For the scales larger than $k = 0.2 \, h \, \text{Mpc}^{-1}$, the results for the various redshift are analogous to the ones of \autoref{fig:signal_to_noise}.
\begin{table}
	\centering
	\caption{Shot noise values used in the computation of the non-linear 21cm power spectrum, at each central redshift $z_c$. For the first redshift bin ($z=0.25$) we assume SKA-Mid Band 2 specifications, while we use SKA-Mid Band 1 parameters for the other bin. We refer to \autoref{tab:SKA_specifics} for more instrumental details.}
	\label{tab:shot_noise}
	\begin{tabular}{clccc} 
	\hline 
    $z_c$  & central redshift & 0.25 & 0.75 \\\hline
$P_{\rm SN}$ & shot noise  &0.72 & 2.4 & $[\mathrm{mK^2}\, h^{-3}\mathrm{ Mpc^3 }]$\\
$N$ & number of data points & 83 & 58 \\
		\hline
	\end{tabular}
\end{table}

With this non-linear data set we perform an analysis similar to the one discussed in the previous sections. We study the constraining power of non-linear 21cm observations alone and combined with CMB. We first assume perfect knowledge on the quantities $\Omega_{\rm HI}$ and $b_{\rm HI}$ linked to baryon physics. Note that, for this ideal case without any nuisance parameters, we assume also that the level of the shot noise is known.

Results are presented in Tab.~\ref{tab:constraints_P21_alone} and \autoref{fig:full_P21}. 

\medskip
Even if only the first two bins are concerned,
the extension of the data set to non-linear scales significantly improves on the constraining power of 21cm observations. We find
\begin{equation}
H_0 = 67.28 \pm 0.33\,\, \text{ km s}^{-1}\text{Mpc}^{-1}\qquad {\scriptstyle(68\%,\ \Hat{P}_0^{\rm NL} + \Hat{P}_2^{\rm NL})}
\end{equation}
thus a relative error of $\sigma_{H_0} = 0.49\%$, competitive with the one from Planck 2018 data alone (i.e. $\sigma_{H_0} = 0.79\%$).

We then test the more realistic case where we vary the nuisances parameters. Along with the eight nuisances for the redshift evolution of $\Bar{T}_{\rm b}b_{\mathrm{HI}}\sigma_8(z)$ and $\Bar{T}_{\rm b}f\sigma_8(z)$ (see \secref{sec:res:nuisances}), we include six additional parameters $P_{\rm SN, i}$, with $i=\{1,\dots,6\}$, to model the shot noise in each redshift bin.

For example, we get
\begin{equation}
    H_0 = 68.3^{+3.4}_{-4.7}\,\, \text{ km s}^{-1}\text{Mpc}^{-1}  \quad {\scriptstyle(68\%,\ \Hat{P}_0^{\rm NL} + \Hat{P}_2^{\rm NL} + \text{ nuis. } (P_{\rm SN}))}.
\end{equation}
The relative error is $\sigma_{H_0} = 5.9\%$, better than the 
corresponding result for the linear-scale case discussed in \secref{sec:res:nuisances} (i.e. $\sigma_{H_0} = 7.4\%$). When we open to nuisances with the non-linear data set, the presence of the shot noise terms helps to fix the height of the power spectrum and we do not loose all the constraining power on $A_s$ and $\sigma_8$. We can thus constrain all the cosmological parameters, with the exception of $\tau$. The posteriors for the shot noise parameters $P_{\rm SN, i}$ and the other nuisances are shown in \autoref{fig:full_nuisances_shot} and commented in \appref{sec:app_nuisances}.

In summary, extending the data set to non-linear has an essential role in increasing the constraining power of the 21cm multipoles alone. Our results suggest that competitive constraints independent from other probes could be obtained with 21cm intensity mapping observations at lower redshifts and non-linear scales. 

\medskip 
When combining the non-linear data set $\Hat{P}_0^{\rm NL} + \Hat{P}_2^{\rm NL}$ $+$ nuisances with Planck data, instead, we do not observe substantial changes in the constraints with respect to the Planck + $\Hat{P}_0 + \Hat{P}_2$ $+$ nuisances case (shown in \autoref{fig:full_results}).
As discussed in \secref{sec:res_P21_P21+Planck}, the improvement in combining the two probes mainly comes from the interaction of opposite degeneracy direction for some of the cosmological parameters between the CMB and the 21cm power spectrum. These are unaffected by the extension to non-linear scales and thus,  when combined with the Planck data, this extended mock data set does not add much information with respect to the linear one.

\section{Conclusions}
\label{sec:conclusions}
In this work, we forecast the constraints on the $\Lambda$CDM cosmological parameters 
for a neutral hydrogen intensity mapping survey with the SKAO telescope, assuming the measurement of the first multipoles of the redshift-space 21cm power spectrum. We construct and analyse this mock data set as an alternative large-scale structure probe alone and in combination with Planck CMB data. 
We model monopole and quadrupole signal of the 21cm power spectrum at linear scales as in \citet{Blake:2019,Cunnington:2020,Soares:2020} and we include in our analysis the full non-diagonal covariance matrix between the multipoles.

\medskip
We follow the SKAO Red Book~\citep{Bacon:2018} proposal and simulate single-dish observations with the SKA-Mid telescope both in Band 2 (frequency range $0.95-1.75$ GHz) and in Band 1 (frequency range $0.35-1.05$ GHz). 
Assuming a Planck 2018 fiducial cosmology, we construct a tomographic data set of observations within six different redshift bins. 
To test the constraining power on the cosmological parameters of the constructed data set, we implement the computation of the likelihood function for the monopole and the quadrupole, fully integrated with the MCMC sampler \texttt{CosmoMC}.
We include a discussion on the impact of our lack of knowledge on the baryonic physics involved in the computation of the 21cm power spectrum, as nuisance parameters in the analysis.

\medskip
We first focus on the 21cm power spectrum measurements at linear scales, that are the preferred target of single-dish intensity mapping observations with SKA-Mid, due to the large beam on the sky. 
However, for the lowest redshifts, the telescope beam is small enough to allow to probe also the non-linear scales. We thus extend our mock data set to non-linearities and the shot noise contribution to check if this could improve on the constraining power.
The results of our analysis can be summarized as follows. 

\medskip
We find that the mock SKA-Mid 21cm observations have a good constraining power on the cosmological parameters. The constraints we obtain are comparable with other probes. E.g., with the 21cm monopole and quadrupole combined, both $H_0$ and $\sigma_8$ are constrained at the $\sim 7\%$ level. The 2D contours presents very marked degeneracies between the parameters, especially in the $\Omega_ch^2$ - $H_0$ and $\ln (10^{10} A_s)$ - $\sigma_8$ planes. 

\medskip
Adding the mock 21cm observations to Planck 2018 CMB data, it is possible to significantly narrow the constraints, with respect to Planck alone.
Although the effect is observable on all the parameters, we get the most significant improvement on $\Omega_c h^2$ and $H_0$, for which the errors are lessen by a fourth. With 21cm multipoles $+$ Planck we estimate $\Omega_c h^2$ and $H_0$ at the $0.25\%$ and $0.16\%$ levels respectively, to be compared with $0.99\%$ and $0.79\%$, obtained with Planck alone. For $\ln (10^{10} A_s)$ and $\sigma_8$ the errors are reduced by more then a factor two. We constrain $(10^{10} A_s)$ at the $0.17\%$ and $\sigma_8$ at the $0.26\%$ level, to be compared with the $0.46\%$ and $0.73\%$ Planck estimates, respectively. Furthermore, we observe that combining the tomographic 21cm data set with CMB alleviates some of the degeneracies between the parameters, resulting in improved constraints. The strongest effect is visible in the $\Omega_c h^2$ - $H_0$ plane. Although 21cm observations are not sensitive to $\tau$, we find that with Planck the improvement on the other parameters is reflected also on $\tau$, reducing the error by a factor two. 

\medskip
To take into account the lack of knowledge on the brightness temperature $T_b$ (that depends on the total HI density $\Omega_{\rm HI}$) and the HI bias $b_{\rm HI}$, 
we repeat our analysis including nuisance parameters. In particular, we consider the combinations $\bar{T}_{\rm b} b_{\rm HI} \sigma_8$ and $\bar{T}_{\rm b} f \sigma_8$, where $f$ is the growth factor. We find that, when we open the parameter space to these nuisances, the constraining power of 21cm multipoles on $A_s$, and consequently on $\sigma_8$, is crucially reduced. However, the results obtained for $\Omega_ch^2$ and $H_0$ remain unaffected, for both the 21cm data set alone and combined with Planck. This result confirms the strength of 21cm tomographic measurements and motivates even more the current observational effort in this field.

\medskip
When we extend the 21cm data set to non-linear scales we find a tightening in the constraints.
The most noteworthy result is that the constraining power of 21cm multipoles observations on $A_s$ and $\sigma_8$ is remarkably improved, even when we open up the parameter space to the nuisance parameters. This is due to fact the information at lower scales helps fixing the amplitude of the power spectrum. 

\medskip
We conclude that 21cm SKAO observations will provide a competitive cosmological probe, complementary to CMB and, thus, pivotal for gaining statistical significance on the cosmological parameters constraints.

\medskip
The formalism presented in this work and the mock data set we construct can be straightforwardly adapted to forecast constraints on the neutrino mass and beyond $\Lambda$CDM models. These extensions are currently under study.
Note that our modelling does not include possible residual foregrounds contamination. A discussion on how the constraints on the cosmological parameters could be biased by this systematic is left for future work. 

\section*{Acknowledgements}

The authors would like to thank Paula S. Soares, Steven Cunnington, and Alkistis Pourtsidou for useful discussion and feedback.
MB and MV are supported by  the INFN INDARK PD51 grant. MV acknowledges contribution from the agreement ASI-INAF n.2017-14-H.0.
MS acknowledges support from the AstroSignals Synergia grant CRSII5\_193826 from the Swiss National Science Foundation.

%%%%%%%%%%%%%%%%%%%%%%%%%%%%%%%%%%%%%%%%%%%%%%%%%%
\section*{Data Availability}

Access to the original code is available upon reasonable request to the corresponding author.

%%%%%%%%%%%%%%%%%%%% REFERENCES %%%%%%%%%%%%%%%%%%

\bibliographystyle{mnras}
\bibliography{Bibliography} 

%%%%%%%%%%%%%%%%%%%%%%%%%%%%%%%%%%%%%%%%%%%%%%%%%%

%%%%%%%%%%%%%%%%% APPENDICES %%%%%%%%%%%%%%%%%%%%%

\appendix

\section{Analytical computation of the monopole and the quadrupole}
\label{sec:app_equations}
In this section we compute the first two coefficient of the Legendre polynomial expansion of \pto. We start from \autoref{eq:ell_P}, i.e.
\begin{equation}
    \Hat{P}_\ell (z,k) = \frac{(2\ell + 1)}{2} \int_{-1}^{1} {\rm d}\mu\, \mathcal{L}_\ell(\mu) \Hat{P}_{21}(z,k,\mu)
\end{equation}
that, substituting \autoref{eq:P_21_full}, becomes
\begin{align}
\begin{split}
    \Hat{P}_\ell (z,k) &= \frac{(2\ell + 1)}{2} \Bar{T}_{\rm b}^2(z) P_{\rm m}(z,k)\int_{-1}^{1} {\rm d}\mu\, \mathcal{L}_\ell(\mu) \tilde{B}^2 (z,k,\mu) \cdot \\
    &\quad\cdot\left[ b_{\mathrm{HI}}(z) + f(z)\, \mu^2\right]^2 \\
    &= \frac{(2\ell + 1)}{2} \Bar{T}_{\rm b}^2\, P_{\rm m}\int_{-1}^{1} {\rm d}\mu\, \mathcal{L}_\ell(\mu)\, e^{ -k^2 R_{\rm beam}^2 (1-\mu^2)}\cdot\\
    &\quad\cdot\left[ b_{\mathrm{HI}} + f\, \mu^2\right]^2 \\
    &= \frac{(2\ell + 1)}{2} \Bar{T}_{\rm b}^2\, P_{\rm m}\, e^{-A} \int_{-1}^{1} {\rm d}\mu\, \mathcal{L}_\ell(\mu)\, e^{ A \mu^2}\left[ b_{\mathrm{HI}} + f\, \mu^2\right]^2 
\end{split}
\end{align}
where we defined $A=k^2R_{\rm beam}^2$ and dropped the explicit dependencies on $z$ and $k$ for the sake of notation. \\

\noindent
{\textbf{Computing} $\mathbf{\Hat{P}_0}$} 

\noindent
Using $\mathcal{L}_0(\mu) =1 $ we obtain
\begin{align}
\begin{split}
\Hat{P}_0 &= \frac{\bar{T}^2_{\rm b}\,P_{\rm m} }{2} e^{-A}  \int_{-1}^{1} {\rm d}\mu\, e^{A\mu^2} \left( b_{\rm HI}^2 + 2 b_{\rm HI} f\mu^2 + f^2 \mu^4\right).
\end{split}
\end{align}
The computation reduces to the following integrals
\begin{align}
\label{eq:app_integrals}
    \begin{split}
        &\int_{-1}^{1} {\rm d}\mu \,\,e^{A\mu^2} = \frac{\sqrt{\pi} \,{\rm erfi}(\sqrt{A})}{\sqrt{A}}, \\
        &\int_{-1}^{1} {\rm d}\mu \,\,e^{A\mu^2} \mu^2 = \frac{ e^A}{A} - \frac{\sqrt{\pi}\,\, {\rm erfi}(\sqrt{A})}{2 A^{3/2}}, \\
        &\int_{-1}^{1} {\rm d}\mu \,\,e^{A\mu^2} \mu^4 = \frac{3 \sqrt{\pi}\,\, {\rm erfi}(\sqrt{A})}{4 A^{5/2}} + \frac{e^A (2 A - 3)}{2 A^2}, \\
    \end{split}
\end{align}
where ${\rm erfi (x)}$ is the imaginary error function.
Thus, we get the following final expression for $\Hat{P}_0$
\begin{align}
\begin{split}
     \Hat{P}_0 &= \frac{\bar{T}^2_{\rm b}\,P_{\rm m}}{2} e^{-A} \Bigg[ b_{\rm HI}^2 \frac{\sqrt{\pi} \,{\rm erfi}(\sqrt{A})}{\sqrt{A}} + 2 b_{\rm HI} f \left(\frac{ e^A}{A} - \frac{\sqrt{\pi}\,\, {\rm erfi}(\sqrt{A})}{2 A^{3/2}} \right) + \\
    &\quad + f^2\left( \frac{3 \sqrt{\pi}\,\, {\rm erfi}(\sqrt{A})}{4 A^{5/2}} + \frac{e^A (2 A - 3)}{2 A^2}\right) \Bigg].
\end{split}
\end{align}

\noindent
{\textbf{Computing} $\mathbf{\Hat{P}_2}$} 

\noindent
Using $\mathcal{L}_2(\mu)=\frac{3\mu^2}{2} - \frac{1}{2}$ we obtain
\begin{align}
    \begin{split}
        \Hat{P}_2 &= \frac{5}{2} \Bar{T}_{\rm b}^2\, P_{\rm m}\, e^{-A} \int_{-1}^{1} {\rm d}\mu\, \left(\frac{3\mu^2}{2} - \frac{1}{2}\right)\, e^{ A \mu^2}\left[ b_{\mathrm{HI}} + f\, \mu^2\right]^2 \\
        &= \frac{5}{2} \Bar{T}_{\rm b}^2\, P_{\rm m}\, e^{-A} \int_{-1}^{1} {\rm d}\mu\, \frac{3\mu^2}{2}\, e^{ A \mu^2}\left[ b_{\mathrm{HI}} + f\, \mu^2\right]^2 - \frac{5}{2}\Hat{P}_0 .
    \end{split}
\end{align}
Adding to the set of \autoref{eq:app_integrals} the integral
\begin{equation}
    \int_{-1}^{1} {\rm d}\mu \,\,e^{A\mu^2} \mu^6 = -\frac{15 \sqrt{\pi}\,\, {\rm erfi}(\sqrt{A})}{8 A^{7/2}} + \frac{e^A (15 - 10 A + 4 A^2)}{4 A^3},
\end{equation}
we can compute the final expression
\begin{align}
\begin{split}
    \Hat{P}_2 &= \frac{15\bar{T}^2_{\rm b}\,P_{\rm m}}{4} e^{-A} \Bigg[ b_{\rm HI}^2 \left(\frac{ e^A}{A} - \frac{\sqrt{\pi}\,\, {\rm erfi}(\sqrt{A})}{2 A^{3/2}} \right) + 2 b_{\rm HI} f \cdot \\
    &\quad\cdot \Bigg( \frac{3 \sqrt{\pi}\,\, {\rm erfi}(\sqrt{A})}{4 A^{5/2}} +  \frac{e^A (2 A - 3)}{2 A^2}\Bigg) +f^2 \Bigg(-\frac{15 \sqrt{\pi}\,\, {\rm erfi}(\sqrt{A})}{8 A^{7/2}} + \\
    &\quad+\frac{e^A (15 - 10 A + 4 A^2)}{4 A^3}\Bigg) \Bigg] - \frac{5}{2}\Hat{P}_0.
\end{split}
\end{align}

\noindent
\section{Nuisances}\label{sec:app_nuisances}
In this section we present and comment the constraints on the nuisance parameters, discussed in~\secref{sec:res:nuisances} and~\secref{sec:res_non_linear}.

\begin{figure*}
	\includegraphics[width=2\columnwidth]{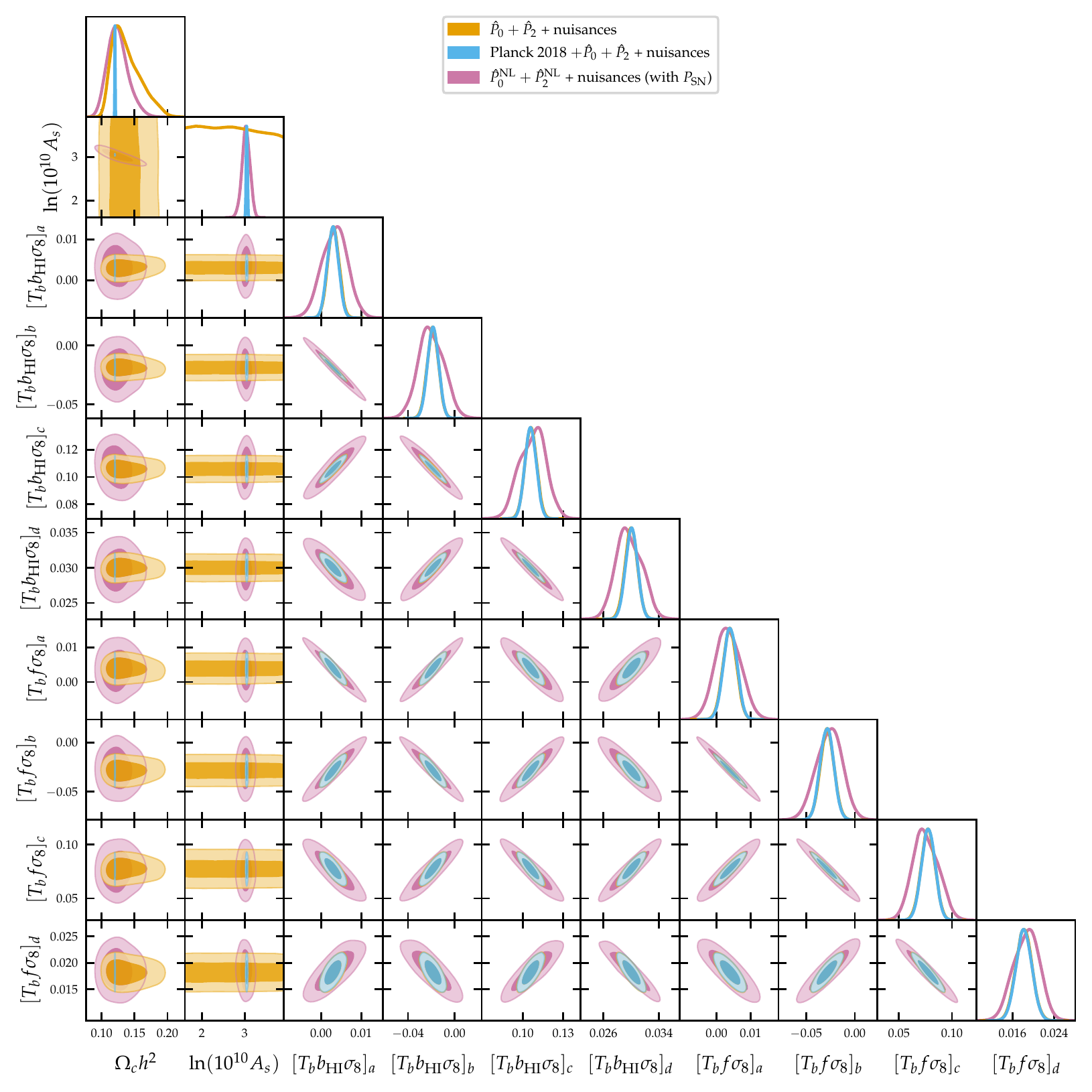}
    \caption{Joint constraints (68\% and 95\% confidence regions) and marginalized posterior distributions on cosmological and nuisance parameters. Here, the label "Planck 2018" stands for TT, TE, EE + lowE + lensing, while the label "$\Hat{P}_0 + \Hat{P}_2$" stands for the baseline tomographic data set for the monopole and the quadrupole combined and with multipole covariance taken into account. The label "$\Hat{P}_0^{\rm NL} + \Hat{P}_2^{\rm NL}$" indicates that the full non-linear data set has been used. The label "nuisances" indicates that we vary the nuisances parameters along with the cosmological ones.}
    \label{fig:full_nuisances}
\end{figure*}

\autoref{fig:full_nuisances} shows the 1D and 2D marginalized posterior distributions of the nuisance parameters $[\bar{T}_{\rm b} b_{\rm HI} \sigma_8]_i$ and $[\bar{T}_{\rm b} f \sigma_8]_i$ ($i=\{a,b,c,d\}$), which describe the redshift evolution of $\bar{T}_{\rm b} b_{\rm HI} \sigma_8(z)$ and $\bar{T}_{\rm b} f \sigma_8(z)$. We consider $\Hat{P}_{21}$ multipoles observations alone and combined with CMB. We find that the nuisances are constrained. The 2D contours show a clear degeneracy with the cosmological parameters, which is eased when we open the parameter space to the shot noise. As discussed above, adding the nuisance parameters we loose all the constraining power on $A_s$, but we recover it when we extend the data set to non-linear scales.

\begin{figure*}
	\includegraphics[width=2\columnwidth]{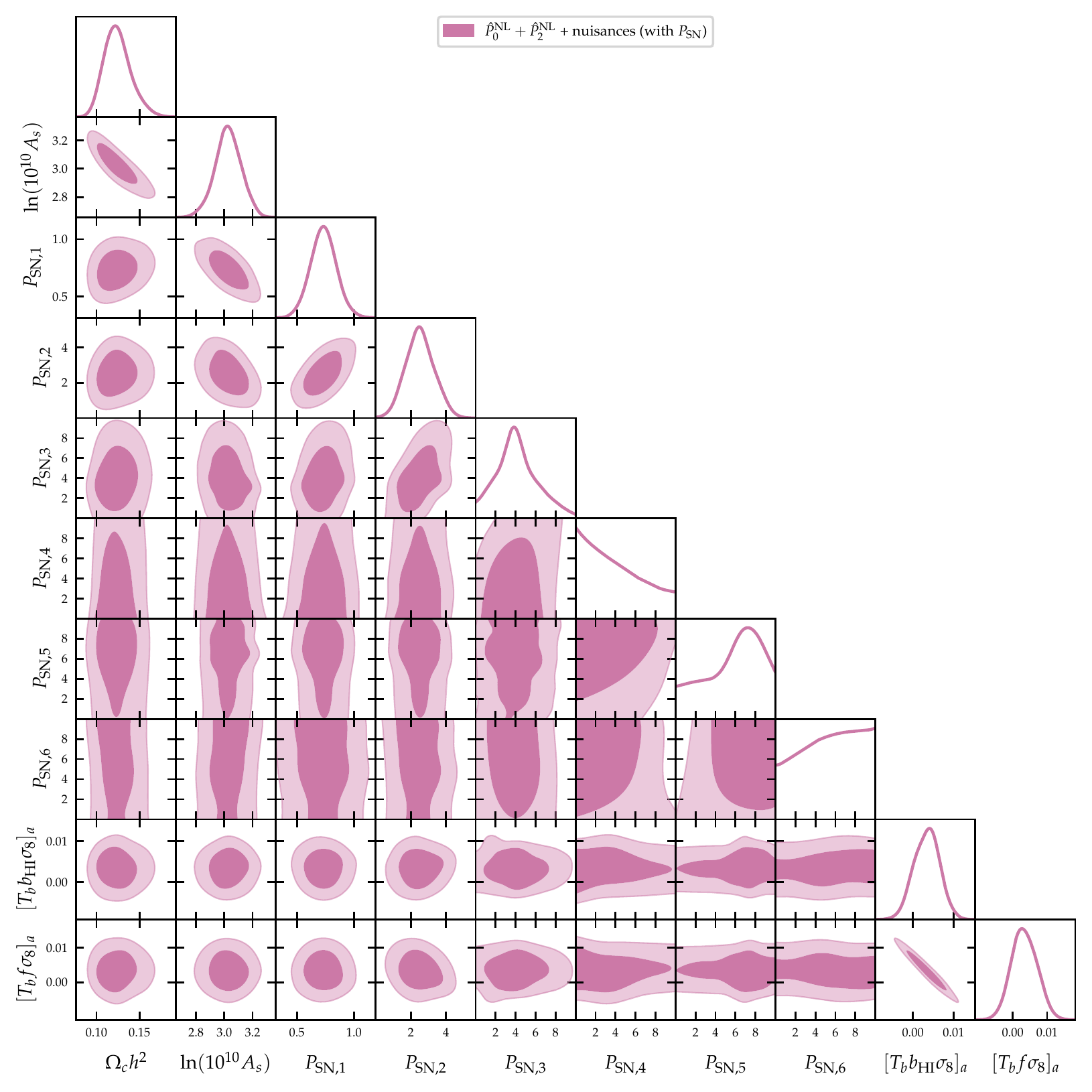}
    \caption{Joint constraints (68\% and 95\% confidence regions) and marginalized posterior distributions on cosmological parameters and the shot noise at different redshifts. Here, the label "$\Hat{P}_0^{\rm NL} + \Hat{P}_2^{\rm NL}$" indicates that the full non-linear data set has been used. The label "nuisances" indicates that we vary the nuisances parameters along with the cosmological ones.}
    \label{fig:full_nuisances_shot}
\end{figure*}

\autoref{fig:full_nuisances_shot}, instead, shows the marginalized posteriors for the shot noise value $P_{\rm SN,i}$ at each redshift bin $i=\{1,\dots,6\}$. We obtain good constraints at low redshift ($i=\{1,2\}$), where we have more data points, while in the highest bins ($i=\{3,4,5\}$), the shot noise results unconstrained. $P_{\rm SN,1}$ and $P_{\rm SN,2}$ present a very mild degeneracy with the cosmological parameters. The shot noise and the other nuisances $\bar{T}_{\rm b} b_{\rm HI} \sigma_8$ and $\bar{T}_{\rm b} f \sigma_8$ are, instead, uncorrelated.

%%%%%%%%%%%%%%%%%%%%%%%%%%%%%%%%%%%%%%%%%%%%%%%%%%

% Don't change these lines
\bsp	% typesetting comment
\label{lastpage}
\end{document}